\documentclass[pre,twocolumn,floatfix,superscriptaddress,showpacs]{revtex4-1}
\usepackage{amsmath,amssymb,latexsym} 
\usepackage{rotating,lipsum}

\usepackage{graphicx}% Include figure files
\usepackage{color}

\begin{document}

\newcommand{\udem}{D\'{e}partement de Physique, Universit\'{e}
de Montr\'{e}al, Montr\'{e}al, Qu\'{e}bec, Canada H3C~3J7}
\newcommand{\nrc}{National Research Council of Canada, Ottawa, Ontario.
 Canada K1A 0R6}

\title
{X-ray Thomson scattering studies on spin-singlet  stabilization 
of highly compressed \\ H-like Be ions
heated to two million
degrees Kelvin.
}

\author
{
 M.W.C. Dharma-wardana}
\email[Email address:\ ]{chandre.dharma@yahoo.ca}
\affiliation{\nrc}\affiliation{\udem}
\author{Dennis D. Klug}
\email[Email address:\ ]{Dennis.Klug@nrc-cnrc.gc.ca}
\affiliation{\nrc}

\begin{abstract}
Experiments at the US National Ignition Facility (NIF)
[D\"{o}ppner et al, Nature {\bf 618}, 270-275 (2023)] have created
highly compressed hot hydrogen-like Be plasmas.
Published analyses of the the NIF experiment have used
finite-$T$ multi-atom density-functional theory (DFT) with
Molecular dynamics (MD), and Path-Integral Monte Carlo (PIMC)
simulations. These methods are very expensive to implement and often lack
physical transparency. Here we (i) relate their results to simpler
first-principles average-atom results, (ii) establish the
 feasibility of rapid data analysis, with good accuracy
 and gain in physical transparency, and (iii) show that
 the NIF experiment reveals high-$T$ spin-singlet pairing of
 hydrogen-like Be ions with near neighbors. Our analysis predicts
 such stabilization over a wide range of compressed densities
for temperatures close to two million Kelvin. Calculations of structure
 factors $S(k)$ for electrons or ions, the Raleigh
 weight and other quantities of interest to X-ray Thomson scattering
are presented. We find that the  NIF data at the scattering wavevector $k_{sc}$ of
 7.89 \AA$^{-1}$ are more consistent with a  density of  
$20\pm2$ g/cm$^3$, mean ionization $\bar{Z}=$3.25, at a temperature of $\simeq$
 1,800,000 K than the 34 g/cm$^3, \bar{Z}=3.4$ proposed by the NIF team.    
The relevance of ion-electron coupled-modes in studying
 small $k_{sc}$ data is indicated.
\end{abstract}

%\vspace{0.5in}
%
\maketitle

\section{Introduction} Matter under extreme densities $\bar{\rho}$,
 temperatures $T$  and pressures $P$
occurs naturally in planetary interiors and astrophysical
 objects. They also occur as transient states that have to be probed
 on sub-nanosecond
 timescales~\cite{Ng05,Poole24,Drake2018,Betti2016,GaffneyHDP18,McBride-Si-19} via
cutting-edge experiments on high-energy-density materials relevant
to experimental astrophysics~\cite{Drake2018}, 
fusion physics~\cite{Betti2016,GaffneyHDP18}
and even for nuclear-stockpile stewardship requirements. The Fermi energies
 $E_F$ of compressed matter are large, and although $T$ (where we use energy units)
 may be nominally high, $T/E_F$ is small and hence
 the name ``warm-dense matter'' (WDM) has
 been used for such highly-correlated energy-dense matter.
 X-ray Thomson scattering (XRTS)~\cite{Gregori03,GlenRed09} 
(using kilo-eV X-rays that penetrate dense matter)  provides
information on the  microscopic quantum states and thermodynamic states of these WDM
samples. For instance, while compressed hot
 silicon~\cite{Poole24,cdwSi20,McBride-Si-19,cdw-SiXRTS24}
is of geophysical interest, highly compressed H, C~\cite{driver12,Militzer2015} and Be
are of interest in  inertial fusion experiments and in astrophysics.
Meanwhile, WDM Aluminum has continued to be a benchmark material whose
EOS and even the dynamic structure factor have been the topic of many studies~\cite
{Ng05,Harb-DSF2018}.

In this study we re-examine the
recent XRTS experiment on highly compressed Be (7.5-35 g/cm$^3$) at temperatures
 close two million Kelvin ($T\sim$ 150-160 eV) conducted at the US national ignition
 facility (NIF)~\cite{Doppner23}. Density functional theory (DFT) using $N$ atoms, 
with $N\sim 64-256$ coupled to molecular-dynamics (MD) optimization of the
ionic structure is one of the tools used in analyzing such XRTS
 data~\cite{SouzaXRTS2014,Plage-XRTS15,xrt-Harb16,CDW-Pool25} where, traditionally,
an evaluation of the mean ionization $\bar{Z}$ is needed~\cite{BethkenZbar20}. This method,
 variously referred to as DFT-MD, KS-MD (Kohn-Sham MD)
and QMD (quantum molecular-dynamics) ~\cite{GaffneyHDP18} will be referred to here
 as QMD due to its brevity and wider usage.
 Finite-$T$ applications of QMD for $T>E_F$, where $E_F$ is the Fermi energy,
are prohibitively expensive for most
 laboratories as the basis sets needed for the
DFT calculations increase rapidly with $T$. Other methods, e.g.,
path-integral Monte Carlo (PIMC) simulations~\cite{DornheimPIMC24,DornheimRw24} 
are even more
 demanding for computer resources.
 D\"{o}ppner et al~\cite{Doppner23} used QMD as well as average-atom (AA)
methods and analyzed their XRTs data at $k=7.9$ \AA$^{-1}$.
 They  concluded that the Be WDM  created at the NIF had a density $\bar{\rho}=34\pm4$
 g/cm$^3$ at $T$= 160$\pm 20$ eV, with a mean ionization $\bar{Z}=3.4\pm 0.1$. 

While a major objective of our study is to re-analyze the XRTS data
 of the NIF experiment and draw attention to the spin-singlet 
stabilization of the hydrogen-like
state of Be atoms, an equally important objective is
to demonstrate that such analyses can be done rapidly and economically
 using a full first-principles approach~\cite{DWP82,Pe-Be,eos95}
 while achieving an accuracy at least comparable to the best available
published results using QMD, PIMC etc., on the NIF-Be study. This is important since
the computationally expensive QMD and PIMC resources needed to analyze
 extreme states of matter used by the NIF team or  by Dornheim et al
are not available to most laboratories.

Average-atom models
construct ions containing a set of $n_b$ bound electrons,
and carrying a mean charge $\bar{Z}$, while the nuclear charge $Z_{nu}=\bar{Z}+n_b$.
Another objective, taken up in this paper is that $\bar{Z}$ is a rigorous DFT concept. 
If AA models are based strictly on DFT, then a one-atom DFT model can be formulated.
In such models,
{\it both} electron- and ion-  many-body problems are reduced using appropriate
XC-functions for electrons and ions respectively, e.g., as in
the neutral pseudo-atom (NPA) model~\cite{DWP82,Pe-Be,eos95,CDW-Pool25}.
Relating the quantities calculated by such AA models to, say,  PIMC calculations,
and defining equivalent quantities for different approaches require care,
 and this too is addressed in this study where we use the NPA
version of an average-atom. This is also referred to as a ``one-atom'' DFT
approach, in contrast to the ``N-atom'' QMD approach. 
  
 All calculations presented in this paper take mere minutes,
 and used only a HP-Inspiron (2018) elitebook laptop, while a
 VASP~\cite{Vasp-lqdSi,VASP} simulation reported here was done elsewhere.
 Furthermore, the one-atom DFT approach for XRTS~\cite{xrt-Harb16},
 and the use of the classical-map (CM) approach for a more rapid calculation of the
structure factor of quantum electrons~\cite{prl1,Bredow15,PDWXC} adopted here
provide a physical picture that may be lacking in purely numerical simulations. 
For instance, PIMC an QMD do not use the distinction between bound and free
electrons, but the physical picture of the 1$s$-bound electron and its delocalization,
developed within the concept of mean ionization is very useful here.

A PIMC study of the NIF experiment by
Dornheim et al~\cite{DornheimPIMC24,DornheimRw24} also revises the NIF estimate and
proposes $\bar{\rho}=22\pm2$ g/cm$^3$ at a temperature of 155.5 eV.
We conclude, in close agreement with the Dornheim et al,
 that the NIF-Be data are consistent with a  density 
$\bar{\rho}=20\pm2$ g/cm$^3, T=155\pm 5$ eV 
and $\bar{Z}=3.25\pm 0.01$.

\section{The NIF experiment and XRTS analyses}
A hohlraum compression of a Be capsule is achieved in the NIF experiment
using 184 optical laser beams. A further eight laser beams generate $\sim$8.9 KeV
X-rays from a zinc foil. They are used for the
probe beam~\cite{Doppner23}. The scattered intensity at selected angles
(i.e., for scattering vectors $\vec{k}_{sc}$) is measured.
 The ratio of the elastic ($el$) to the inelastic ($inel$)
contributions to the full scattering intensity, denoted by $r(k)=I_{el} /I_{inel}$ at
the wave vector $\vec{k}$, can be directly measured. Since the WDM is in a plasma
state, we assume a uniform fluid and only the magnitude of $\vec{k}$ is 
relevant for the data analysis. 

We use atomic units with $\hbar=m_e=|e|=1$, and
 $T$ will usually be given in energy units of eV (1 eV=11,604K). We also define
the electron- and ion-  Wigner-Seitz radii $r_s=[3/(4\pi\bar{n}]^{1/3}, 
r_{ws}=[3/(4\pi\bar{\rho}]^{1/3}$, where $\bar{n}$ is the average 
free-electron density in the plasma, while $\bar{\rho}$ is the average
ion density. The number of free electrons per ion
(experimentally measured using a Langmuir probe in low-$T$ plasmas) is
$\bar{Z}$, with  $\bar{Z}=\bar{n}/\bar{\rho}$. The Fermi wavevector
$k_F=(2\pi^2\bar{n})^{1/3}$ and the Fermi energy $E_F=k_F^2/2$ 
are important scales of energy and momentum for the electron subsystem.

The temperature of the sample has been estimated~\cite{Dornheim-Tem23}
 from detailed-balance considerations~\cite{GlenRed09} applied
 to the XRTS signal, although more extensive modeling is needed in
 two-temperature plasmas  where the ion temperature $T_i$ differs from the electron
 temperature $T_e$~\cite{xrt-Harb16}. The use of the imaginary-time
correlation function approach for two-temperature systems has been proposed
in Ref.~\cite{Vorberger24}. Several theoretical  analyses of the NIF experiment
~\cite{Doppner23,DornheimPIMC24,Bellenbaum24,DornheimRw24} are available, and
some use the  well-known Chihara decomposition~\cite{Chihara2000} together with QMD
calculations~\cite{Plage-XRTS15} or with the NPA~\cite{xrt-Harb16,CDW-Pool25}.
 
The PIMC~\cite{DornheimPIMC24} simulations do not need the Chihara
decomposition. However, it should be  noted that NPA calculations, as
well as average-atom calculations, unambiguously yield a complete set of
 eigenfunctions
with negative and positive energies $\epsilon_\nu(r)$, for
liquid metals and plasmas (away from metal-nonmetal transitions). Hence
 the separation of the spectrum into a bound part and a free 
part is not an {\it ad hoc} procedure, nor is it dependent on any
``chemical'' models of valance theory, but depends on 
the AA model itself.
This dependence can be pinned down to a unique result if the model
is required to be fully consistent with DFT. There are systematic approaches
available even near metal-nonmetal transitions and resonances, where
partially delocalized electrons have to be accounted for~\cite{hop1992}.
Recent developments in PIMC calculations have enabled the study of such 
regimes (e.g., at metal-insulator transitions)
 {\it ab initio}~\cite{DornheimPhysChemLett24}. Such simulations involve
 thousands of electrons, and are very expensive compared to one-atom DFT
methods.

Dornheim et al introduce
calculations of the Rayleigh weight $R_w(k)$, as well as the
$el/inel$ ratio  $r(k)$ as model-insensitive
 quantities for determining
 $\bar{\rho},T$ and other relevant
information of the WDM state. Calculations of $R_w(k), r(k)$ are
easily available within the NPA approach used here. However, many of the
quantities used in the PIMC analysis, e.g., the electron-ion structure factor
$S_{ei}(k)$, do not correspond to the $S_{ei}(k)$ used in the theory of electrons
and ions that uses ions of mean charge $\bar{Z}$ as one of the components,
even if the $\bar{Z}$ implicit in the PIMC calculation agrees with that of the
AA calculation. In effect,
 PIMC uses a two-component fluid consisting of {\it nuclei} and electrons, and deals with
the electron-nuclear structure factor $S_{en}(k)$ involving all electrons, instead
 of $S_{ei}(k)$ that includes only the free-electron cloud, as defined by
 Chihara~\cite{Chihara2000}. These differences are further discussed below. 

Since the scattered intensity is proportional to the {\it e-e} dynamic
 structure (DSF) factor $S_{ee}(k,\omega)$,
all theoretical approaches reduce to a manipulation and simplification
of $S_{ee}$ together with $S_{ei}$ and the ion-ion structure factor $S_{ii}$.
Since ion-dynamics is not resolved in
most experiments, the frequency-integrated structure factors
$S_{a,b}(k)$,  $a=e,i$ can be used where appropriate. We
define below, frequently used quantities of interest.
These include the elastic and inelastic components of
$S_{ee}(k)$ where the use of an appropriate value of  $\bar{Z}$ is implied
in this separation, as in Chihara's approach~\cite{Chihara2000}. 
\begin{eqnarray}
S_{ee}(k) &=&S_{el}(k)+S_{inel}(k)\\
r(k)&=&S_{el}(k)/S_{inel}(k)\\
F_e(k)&=&\{n_b(k)+n_f(k)\} \\
F_I(k)&=&F_e(k)^2S_{ii}(k)
\end{eqnarray}
Here $F_e(k)$ is the electron form factor made up of the bound- and free-
electron densities (in $k$-space), i.e., $n_b(k), n_f(k)$ respectively, obtained
by Fourier transformation of $n_b(r)$ and $n_f(r)$-$\bar{n}$. Furthermore,
$S_{ii}(k)$ is the ion-ion structure factor. This is the same as the
nuclear-nuclear structure factor $S_{nn}(k)$ obtained from
expensive PIMC simulations of a mixture of, say, 25 Be nuclei
and 100 electrons~\cite{DornheimPIMC24}. The last equation above, for $F_I(k)$,
gives the so-called ``ion-feature'' used in XRTS studies. 

\begin{figure}[t]                    % fig. 1
\includegraphics[width=0.96\columnwidth]{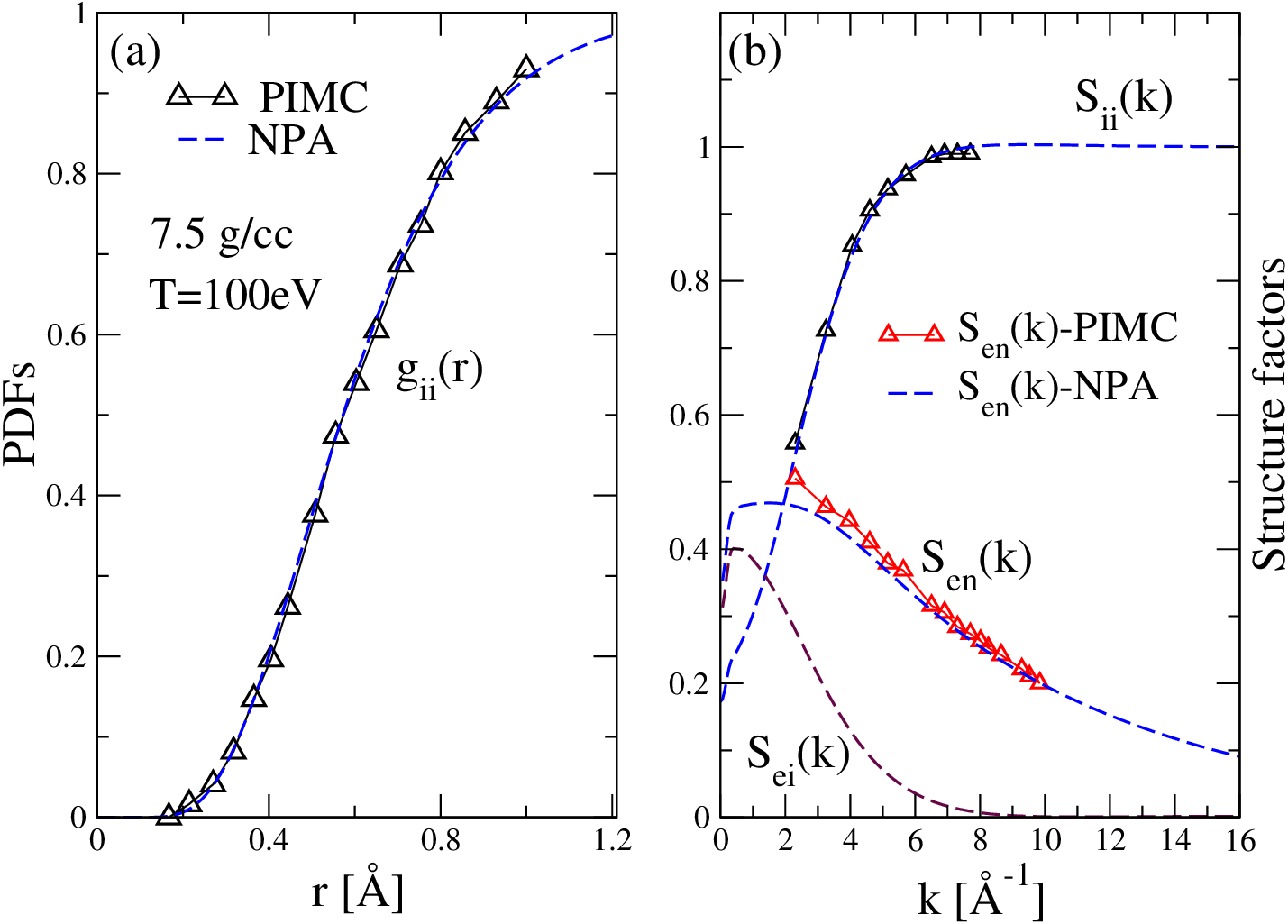}
 \caption{\label{grSk7p5.fig}(online color) (a) The NPA ion-ion
$g(r)$ (dashed line)
for Be at 7.5 g/cm$^3$, $T$=100eV compared with PIMC calculations
 (triangles)~\cite{DornheimPIMC24} (b) The ion-ion structure factor $S_{ii}(k)$,
the electron-ion $S_{ei}(k)$ and the electron-nuclear $S_{en}(k)$ from NPA and PMIC.
}
\end{figure}

In QMD, while the electron-electron many-body
 problem has been reduced to a one-body problem, the ion-ion problem has not
 been reduced in the same manner. Hence in QMD, $N$-atoms are placed
 in a simulation box and the multi-center Kohn-Sham electronic density
 $n(\vec{r},\vec{R}_1,\ldots,\vec{R}_N)$  is calculated for each fixed
 ionic configuration ${\vec{R}_I},I=1$ to $N$,  treated as a realization
 of a periodic crystalline solid. The ionic configurations
 ${\vec{R}_I}$ are evolved using classical MD and a configuration average is
 taken to access physical quantities pertaining to the fluid state. In calculating
 the XRTS signal, or the mean ionization $\bar{Z}$, usually the $N$-atom output
 of QMD has to be decomposed into an  average single-atom contribution,
 as discussed, e.g., in Plageman et al~\cite{Plage-XRTS15}. The
$\bar{Z}$ used in the NIF-study of Be is presumably based on the
method of Bethkenhagen et al.~\cite{BethkenZbar20}.

We use QMD or PIMC only to establish benchmarks.
For the bulk of the calculations we use NPA, a method
that has been used in many studies since 1982, with
details of  practical implementations published in Refs.~\cite{Pe-Be,eos95,cdwSi20},
and validated  by comparing with  QMD, PIMC~\cite{driver12}, and other calculations,
 even for dynamic structure-factor calculations~\cite{Harb-DSF2018}.
We have shown in previous publications that the NPA which uses a single-nucleus-based
DFT reduction of the many-ion and many-electron problem, yields the $g_{ii}(r)$
 and other quantities that  agree very well with those
obtained from QMD and PIMC. Furthermore, in the present study too
we compare with the PIMC results of Dornheim et al, and give QMD results for
 Be using the Vienna Ab-initio Simulation
 Package~\cite{VASP}, as detailed in the appendix.

 In Figure~\ref{grSk7p5.fig} we compare our NPA calculations 
(whose details are given below)
 for Be at 7.5 g/cm$^3$
at 100 eV with PIMC calculations~\cite{DornheimPIMC24}, showing good agreement.
The $S_{en}(k)$-NPA, constructed from the NPA charge densities
agrees with the PIMC $S_{en}(k)$  in the relevant
range of $k$, i.e., $k>k_F$. The small-$k$ region is not accessible via PIMC
due to limitations of simulation-box size.
 An approximate extension of $S_{ei}(k)$
and $S_{en}(k)$ for small-$k (< k_F)$ presented in the figure uses
 an approximate two-component-fluid description discussed below, and in the appendix.

\section{NPA calculations for the NIF-Be experiment}
The details of the NPA calculation are well established~\cite{DWP82,Pe-Be,eos95}.
However, we give a brief review of it in the appendix touching on how the
number of free electrons per ion, viz., $\bar{Z}$ is determined via the Friedel sumrule
and the minimization of the total free energy of the system.

\begin{figure}[t]                    % fig. 2
\includegraphics[width=0.99\columnwidth]{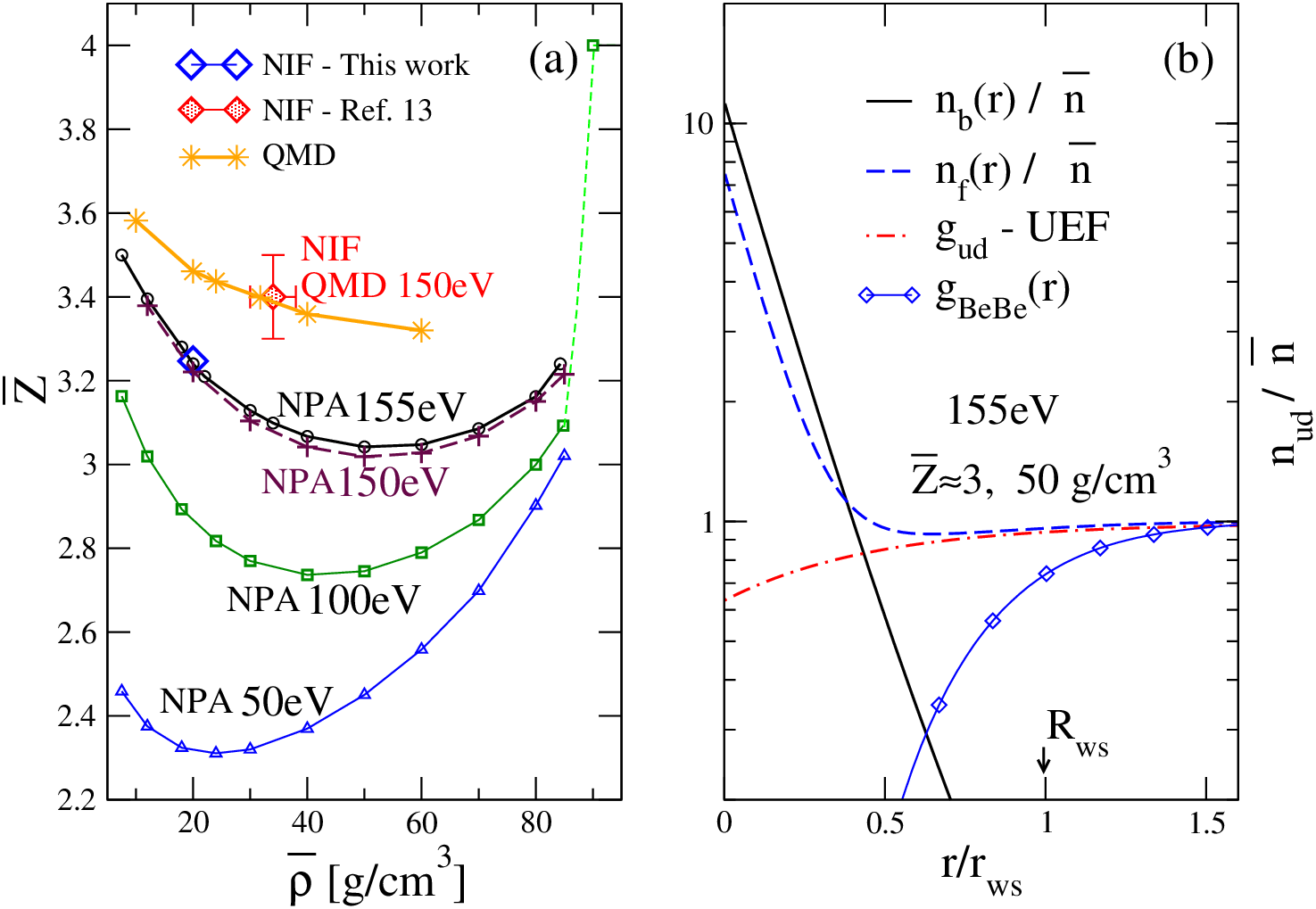}
 \caption{\label{zbargud.fig}(online color) (a) The NPA mean ionization $\bar{Z}$ along
four isotherms. The range of $\bar{\rho}$ with $\bar{Z}\simeq 3$ widens as $T$
 increases. The $\bar{Z}$ and $\bar{\rho}$ estimates for the NIF-Be are indicated
 as `Diamond' data points). The QMD estimates of $\bar{Z}$, Ref.~\cite{Doppner23} 
at 150 eV,(asterisks) are higher than NPA $\bar{Z}$ at 150 eV estimates (+ symbol).
The onset of full ionization to $\bar{Z}=4$ is shown for $T=100$ eV
($\bar{\rho}\sim$ 85-90 g/cm$^3$.  
(b) The data for the case of a single electron (of specific spin) occupying the
 $1s$ at 155 eV, viz., 50 g/cm$^3$ is displayed.
 The $g_{\rm BeBe}(r)$ penetrates the WS-sphere; thus the 1$s$ nearest-neighbour
 eigenstates will have opposite spin states (c.f., exclusion principle). The up-down e-e 
PDF for a uniform electron fluid (UEF) at the same $\bar{n},T$ (dot-dashed line)
shows a weaker up-down enhancement at $r=0$, but merges with $n_f(r)/\bar{n}$
for $r > r_{ws}/2$.}
\end{figure}

\subsection{Mean-ionization isotherms} 
\label{zbarMain.sec}
In Fig.~\ref{zbargud.fig}(a) we display
 the variation of $\bar{Z}$ along isotherms
$T=50,100,150$ and 155 eV. The significantly higher $\bar{Z}$ estimate
of Ref.~\cite{Doppner23} using QMD
for $T=150$ eV is also shown (curve with asterisks).

The reasons for the overestimate of $\bar{Z}$ in current QMD calculations
have been noted and discussed in previous studies as well. DFT is a theory
for the {\it total energy} and is not designed to give physical one-electron
energies, occupations or bandgaps. These quantities (as calculated by most codes)
contain  self-interaction  errors as well as
discontinuities in the XC-functional that arise when an $N+1$ electron system
becomes an $N$-electron system (see Kohn, Ref.~\cite{Kohn86}) interacting with an electron
in a higher Kohn-Sham energy band, within a scheme of non-interacting Kohn-Sham electrons.
These bands are non-physical DFT constructs; thus, normal-density Ge appears
as a metal without a bandgap. As pointed out by
 Dharma-wardana~\cite{cdw-Carbon10E6-21},
the calculations of $\bar{Z}$ by Bethkenhagen et al~\cite{BethkenZbar20}
use DFT eigenstates and eigenvalues that are not
corrected for these self-interaction effects and discontinuities in the XC-functionals.
 This is possibly the main reason for the
difference in the values of $\bar{Z}$ evaluated via the NPA,
and via QMD by Bethkenhagen et al, and via average-atom calculations~\cite{Fauss21}.
 Similarly, counting occupation numbers to obtain
$\bar{Z}$ from Kohn-Sham bound states 
 is not formally justified, as the solution of a
 Dyson equation inclusive of GW-corrections~\cite{PDW-Dyson84} is needed.

Other authors, e.g., Gawne et al~\cite{Gawne24} have also
 discussed ``quantifying ionization in hot dense plasma'' using
 Kohn-Sham eigenstates. Alternative models based on transport properties
have been examined, e.g., in  Sharma et al~\cite{Sharma25} where a
Born effective-charge model for the determination of $\bar{Z}$ is proposed,
 without an {\it ab initio} construction of the underlying atomic model. 
More details of the evaluation of $\bar{Z}$ in the NPA model using the Friedel
 sum rule at the minimum of the total Helmholtz free energy of the electron-ion
system are given in the appendix. Some authors have (erroneously) claimed that
$\bar{Z}$ is not the mean value of any quantum operator and is hence not
admissible in quantum calculations. We give its operator form in Sec. 2 of the Appendix,
and also point out that the temperature $T$ does not have a corresponding
quantum operator in the usual sense.

Although the NPA $\bar{Z}$ had reached close to 3.5 near 7 g/cm$^3$,
compression does not push it to $\bar{Z}=4$. Instead, the  isotherms
 show a counter-intuitive {\it decrease} in ionization on compression,
 reaching towards
more binding and then only proceeding to full ionization ($\bar{Z}=4$) at
 around $\bar{\rho}>80-85$ g/cm$^3$. Compression drives the field ions into the
Wigner-Seitz cell of the central ion, as seen from the $g_{ii}(r)$ displayed
in Fig.~\ref{zbargud.fig}(b). At $T=$155 eV, Be-ions become closest to $\bar{Z}=3$
at about $\bar{\rho}=$50 g/cm$^3$. The only available bound state at 155 eV
 is the 1$s$ state,
 here calculated as the Kohn-Sham state $\phi_{nl}(r),n=1,l=0$. The
corresponding $n_b(r)$ is displayed in Fig.~\ref{zbargud.fig}(b),
 for 50 g/cm$^3$ at 155eV
and corresponds nominally to a single electron with a specific spin,
 say, up-spin in $\phi_{1s}(r)$.
Its density, plotted on a log-$y$ scale is essentially linear.
This density meets the field-ions within finite values of $g(r)$;
 hence the nearest-neighbour $\phi_{1s}$, taken
with an opposite spin would form a linear combination state that
 stabilizes the $\bar{Z}=3$
state. This also delocalizes the 1$s$ bound state from being a pure atomic state,
 into a 
 state {\it partially bound to the ion distribution} as well. At lower temperatures,
e.g., 100 eV, the H-like Be with just one electron in the 1$s$-state,
with $\bar{Z}=3$, occurs at 80 g/cm$^3$.

\subsection{XC-functionals}
As the NPA is an
all electron calculation which optimizes both the electron distribution $n(r)$ 
{\it and} the ion distribution $\rho(r)$ to minimize the Helmholtz free energy,
 it captures the best average-ion picture
 consistent with these electronic
interactions brought in via the Hartree and the finite-$T$ XC-functional.
The e-e XC functional of Perrot and Dharma-wardana (PWD)~\cite{PDWXC} is used
in the NPA implementation. However, unlike PIMC simulations, the current implementation
of the NPA does  not use spin-density functional theory. The PDW e-e XC-functional
 used here~\cite{PDWXC} is based on the classical map for the electron fluid fitted to
calculations in the range $1\le r_s\le 10$ with suitable analytic limits imposed for
the Gellmann-Breuckner and Debye-H\"{u}kel limits. However, the fit to Quantum Monte
 Carlo data at finite-$T$ given by Karasiev et al~\cite{Karas18}, and by Groth
 et al,(GDS)~\cite{GDS17} are expected to be more accurate, especially for $r_s$
 outside the fit range  of the PDW version.  Comparisons of $F^{ee}_{xc}$ and
$V_{xc}$  using the the e-e
 XC functionals of GDS and PDW for the conditions of NIF-Be are given in the Appendix,
and show some differences that do not produces significant differences in the final NPA
calculations. 
 Furthermore, in contrast, the QMD calculations seem to have
simply used the $T=0$ XC-functional of Perdew, Burke and Ernzerhof~\cite{PBE96}. These
suggest that finite-$T$ XC-effects are of negligible importance for the NIF-Be conditions.     
\section{Pair distribution functions and Structure factors}
The compressed Be fluid with only 1$s$ occupation may be viewed as an electron fluid
with strongly enhanced up-down e-e correlations. The e-e $g_{ud}(r)$ for a uniform
 electron fluid (UEF) at the same $T,\bar{n}$ as the Be-WDM at 50 g/cm$^3, T=$155 eV,
 calculated using the classical-map approach~\cite{prl1} is also shown in
 Fig.~\ref{zbargud.fig}(b).  The $n_b(r)/\bar{n}$ approximates to $g_{ud}(r)$ of
 the Be-WDM only for  large-$r>r_{ws}$. It should be emphasized that the NPA
calculation is independent of the classical-map calculation. The latter uses the
value of $\bar{Z}$ obtained from the NPA calculation to set the density parameter $r_s$
used in the classical-map calculation.     

The electron densities $n(r)=n_b(r)+n_f(r)$ for $T=155$ eV, $\bar{\rho}=20$ g/cm$^3$
are relevant to our NIF-data analysis. They provides the densities used for NPA estimate
 of $g_{ei}(r)$ and $S_{ei}(k)$.
Since the NPA has provided the Kohn-Sham one-electron states $\phi_{\nu}, \nu=n,l$ for
bound states, or $k,l$ for continuum states, they can be used
 to construct an electron-electron response function
from which the e-e structure factor $S_{ee}(k)$ as modified by the ionic interactions
could be obtained~\cite{DW-unpub}. Thus all three
static structure factors are in principle at our disposal from the NPA calculation.
 However, a more simplified  calculation
of $S_{ee}$ inclusive of non-local XC-interactions can be obtained using the classical
 map (CM) of the electron-fluid~\cite{prl1,PDWXC},  using the $\bar{Z}$ obtained from the
 NPA to construct the CM. More details are given in the appendix.

\begin{figure}[t]                    % fig. 3
\includegraphics[width=0.95\columnwidth]{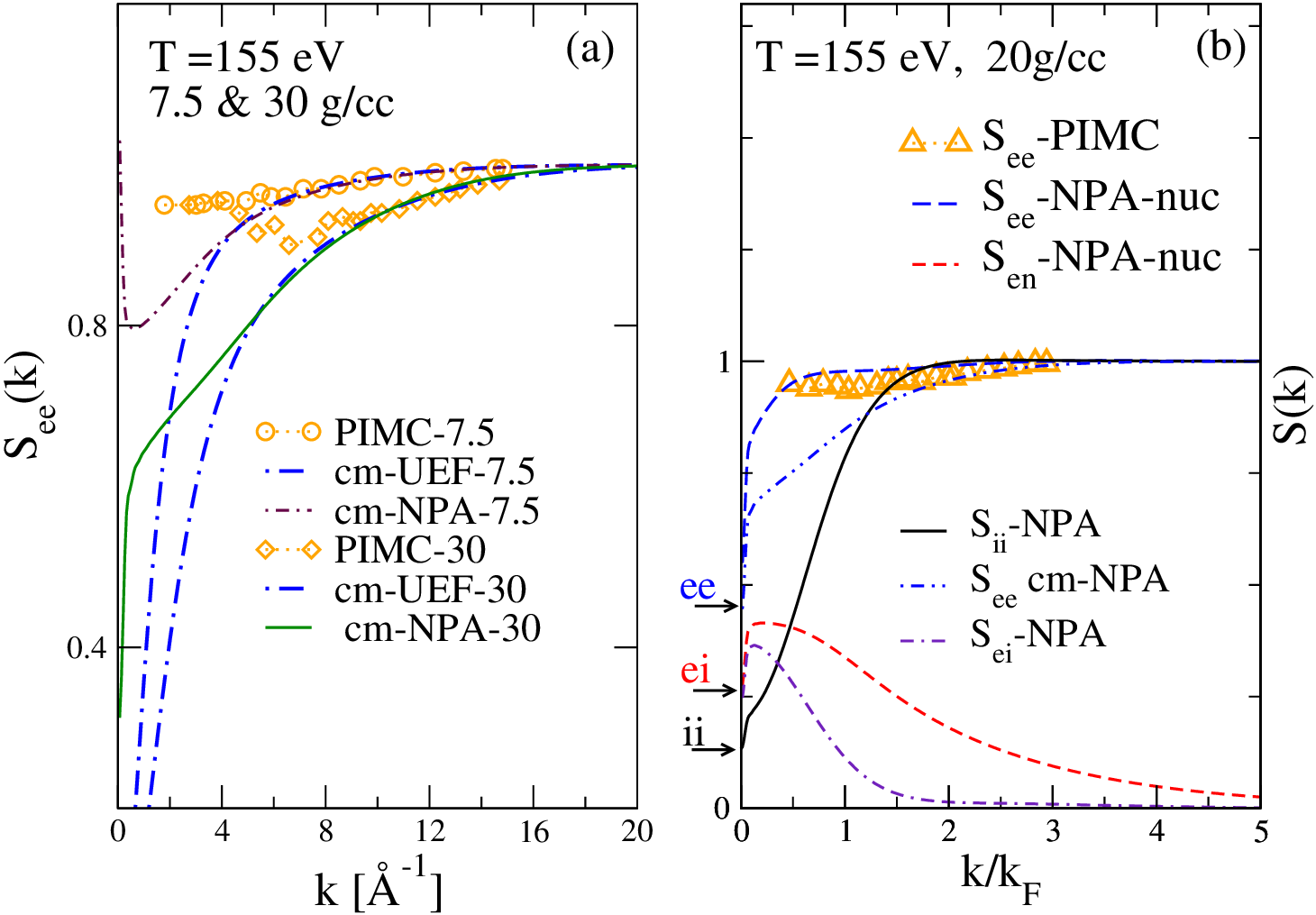}
 \caption{\label{Sk2panel.fig}(online color) (a) The $S_{ee}(k)$ from
 PIMC, uniform-electron fluid (UEF), and from Eq.~\ref{s_eeSum.eqn}
 for two densities, viz, 7.5 g.cm$^3$ and
 30 g/cm$^3$. The PIMC data are from ref.~\cite{DornheimPIMC24}.
(b) The PIMC $S_{ee}$ matched with $S_{ee}$-NPA-nuc constructed
using Eq.~\ref{S_ee-pimc.eqn}.
for Be at 20 g/cm$^3$ and $T=155$ eV.  
}
\end{figure}

We note that the PIMC $S_{ei}(k)$ does not extend to $k<2$ \AA, presumably due to the 
limited size of the simulation box of the 25-atom sample.
In Fig.~\ref{Sk2panel.fig} we display our NPA and classical-map 
based e-e structure factor $S^{cm}_{ee}(k)$, as well as  NPA results
 for $S_{ii}(k), S_{ei}(k)$, for Be at 20 g/cm$^3$ and $T=155$ eV.
  The classical-map calculation~\cite{prl1} is for the interacting uniform
electron fluid with a rigid uniform positive background; i.e., it is
essentially a one-component calculation. This increasingly
 agrees with the
two-component PIMC calculation as $k$ increases, for $k>k_F$. 
The CM-calculation can be
approximately extended to a two-component form (c.f. Chihara, Eq.38
 of Ref.~\cite{Chihara2000})
using the $n_f(k)$, and $S_{ii}(k)$ from the NPA calculation.
 As such, the extended $S_{ee}(k)$ are labeled
as cm-NPA-$\bar{\rho}$ for each density $\bar{\rho}$ in the figure.
\begin{eqnarray}
\label{s_eeSum.eqn}
S_{ee}(k)&=&S^{cm}_{ee}(k)+S_{ii}(k)|n_f(k)|^2/\bar{Z}\\
S_{ei}(k)&=&n_f(k)S_{ii}(k)/\surd{\bar{Z}}
\end{eqnarray}  
This Chihara extension obeys the compressibility sumrule. That is, if $\xi$ is
the isothermal compressibility of the electron-ion fluid, then:
\begin{equation}
\bar{\rho}T\xi=S_{ii}(0)=S_{ei}(0)/\surd{\bar{Z}}=S_{ee}(0)/\bar{Z}.
\end{equation}
Chihara's extension is an approximate model and not the result of a full two-component
 calculation. The latter is feasible with the classical-map hyper-netted chain
 (CHNC) procedure~\cite{prl1}. However, we have left this for a future study
since the CHNC procedure itself has
an ambiguity in assigning a temperature for the electron-ion interaction~\cite{Bredow15}.
No such ambiguity exists in the NPA calculations.

 It is seen that the Chihara extension of $S^{cm}_{ee}(k)$
does not fully capture the $S_{ee}(k)$ from PIMC. 
In panel (b) of  Fig.~\ref{Sk2panel.fig}
we display an extension of $S^{cm}_{ee}(k)$ which is more successful in
capturing the PIMC-$S_{ee}(k)$. Here again we note that the PIMC is for a two-component
mixture of electrons and {\it nuclei}. Hence we consider the forms:
\begin{eqnarray}
\label{S_ee-pimc.eqn}
S_{ee}(k)&=&S^{cm}_{ee}+S_{nn}(k)|n_f(k)+n_b(k)|^2/Z_{nu}\\
S_{en}(k)&=&S_{nn}(k)|n_f(k)+n_b(k)|/\surd{Z_{nu}}\\
S_{nn}(k)&=&S_{ii}(k)
\end{eqnarray}
Here $S_{nn}(k)$ is the nuclear-nuclear structure factor, and $Z_{nu}$ is the
nuclear charge. The $S_{nn}(k)$ of the PIMC calculation is identical with the
$S_{ii}(k)$ of NPA calculation provided that the $\bar{Z}$ estimate implicit in
the PIMC calculation agrees with that of the NPA.
The  $S_{ee}(k)$ from PIMC and cm-UEF for
 two densities, viz, 7.5 and  30 g/cm$^3$ are displayed. Note the different small-$k$
 behavior of the PIMC $S_{ee}(k)$ and the
CM-$S_{ee}$. The 30 g/cm$^3$ curve follows the classical-map
result up to about 5\AA$^{-1}$, i.e, $k\sim k_F$, and deviates upwards
and gives a small-$k$ behavior similar to that of the 7.5 g/cm$^3$
case. Setting $S_{nn}=S_{ii}(k)$, we can evaluate $S_{ee}(k)$ using
 Eq.~\ref{S_ee-pimc.eqn}
to get the curve labeled ``$S_{ee}$-NPA-nuc'' to compare with the PIMC-$S_ee(k)$.
This Eq.~\ref{S_ee-pimc.eqn} closely approximates the PIMC-$S_{ee}$ and reveals
its physical content and its difference from Eq.~\ref{s_eeSum.eqn}  

The CM-$S_{ee}(k)$ does not include the response of the ion-distribution as
 the UEF assumes a rigid neutralizing
 background. However, if the ion distribution is included as in a two-component
plasma, the major effect is on the small-$k$ region where ion-acoustic coupled
 modes are formed and it affects all the response functions $S_{ab}(k), a,b=e,i$ to
 the appropriate degree. The ion-acoustic coupled modes are of such long wavelength that
 PIMC simulations with 25 Be atoms cannot capture them.
Their effect is to modify the small-$k$ behavior of  $S_{ee}(k)$ and in
fact enforces the compressibility sum rule.
However, these effects do not significantly affect our Rayleigh weight calculation,
since the XRTS scattering $k=7.89$\AA$^{-1}$ in the present case is in the high-$k$
regime.

\begin{figure}[b]                    % fig. 4
\includegraphics[width=0.9\columnwidth]{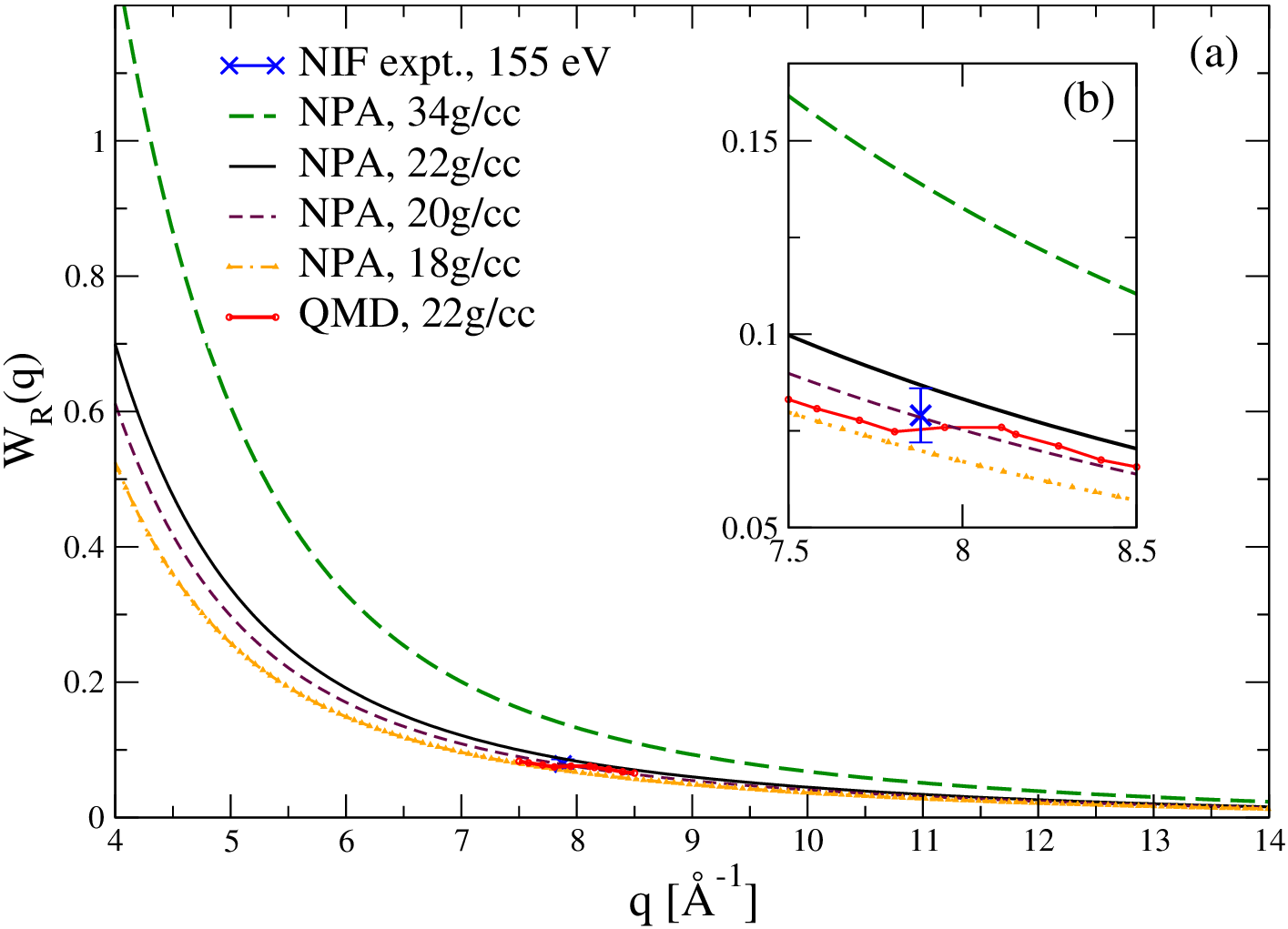}
 \caption{\label{Rw22g155ev.fig}(online color) (a) Rayleigh weight calculations
 for the NIF-XRTS signal
at $k_{sc}$=7.89 \AA$^{-1}$ compared with QMD calculations of
Dornheim et al, see Fig.3 of Ref.~\cite{DornheimRw24}. The inset
(b) gives an expanded view near our NIF-Be estimate.
}
\end{figure}

\section{Calculation of the  Rayleigh weight}
Dormheim et al~\cite{DornheimRw24} propose to use their calculated
 $S_{ei}(k)$ and $S_{ii}(k)$
to calculate the Rayleigh weight $R_{w}$ defined by them as
\begin{equation}
\label{Rw.eqn}
R_w(k)=S_{ei}(k)^2/S_{ii}(k).
\end{equation}
As the PIMC calculations are for a two-component system of electrons and nuclei, as
discussed in the context of Fig.~\ref{Sk2panel.fig}, we note that their $S_{ei}(k)$ is
$S_{en}(k)$ and it includes both bound electrons and free-electrons; hence it is
more appropriately denoted as $S_{en}(k)$. Furthermore, it is {\it not} 
the same as $S_{ei}(k)$ derived from $g_{ei}(r)=n_f(r)/\bar{n}$ normally discussed within
 NPA/AA models as well as in Chihara's theory. The structure factors obtained
from the bound- and free- electron
 densities $n_b(k),n_f(k)$ associated with a single nuclear center as calculated
 by the NPA as well as via AA procedures need to be extended to reflect the two-fluid
structure of the system~\cite{Furutani90}, as briefly discussed in the appendix.
However, for $k>k_F$, the following definition of the Rayleigh weight seems to be
adequate and consistent with the form used by Dornheim et al, viz., Eq.~\ref{Rw.eqn}
 when using quantities calculated via the NPA or AA models.
\begin{equation}
\label{Rw-ion.eqn}
R_w(k)=\{n_b(k)+n_f(k)\}^2/\{\bar{Z}S_{ii}(k)\}.
\end{equation} 
In fig.~\ref{Rw22g155ev.fig} we compare our $R_w(k)$, Eq.~\ref{Rw-ion.eqn} based on an
 ion-electron fluid, with the calculations of Dornheim et al~\cite{DornheimPIMC24} based
 on PIMC and QMD simulations for a nuclei-electron fluid, Eq.~\ref{Rw.eqn}. The figure
shows that the results of the  NPA calculations, obtained
 via methods very different to the PIMC and QMD
calculations provide strong support to one another as they agree and fall well within
the likely error estimates of the two methods.

The scattering wavevector at $k_{sc}$=7.89 \AA$^{-1}$ corresponds to $k/k_F=1.56 $ for
$\bar{\rho}=20.0$ g/cm$^3$ at $T=155$ eV, and $\bar{Z}=3.247$.
It should be noted that while results for smaller scattering vectors, e.g., 
for  $k_{sc}<k_F$ would reveal more information on correlated behavior, these
fall increasingly into the regime where ion-density fluctuations and
electron-density fluctuations form coupled modes (ion-acoustic modes).
Typical plots (Fig. 3, 4 of Ref.~\cite{ELR98}) of the frequency-dependent
 e-e and i-i structure factors 
(Figs. 3, 4) and the relevant theory for quantum systems are given
 in Ref.~\cite{ELR98}. The theory for weakly-coupled classical two-component
electron-ion systems inclusive of coupled-mode formation, and
 including quantum-diffraction effects is given in
Gregori et al~\cite{Gregori03}. 

The study of the small-$k$ scattering regime will increasingly require
 two-component plasma models
constructed using the outputs of the usual NPA or AA calculations.
 Similarly, PIMC calculations and QMD will need very-large simulations
 consistent with the long-wavelength nature of ion-acoustic modes if 
small-$k$ scattering data are to be treated.

\section{Transport properties}
\label{sigkappa.sec}
\begin{figure}[t]                    % fig. 5
\includegraphics[width=0.96\columnwidth]{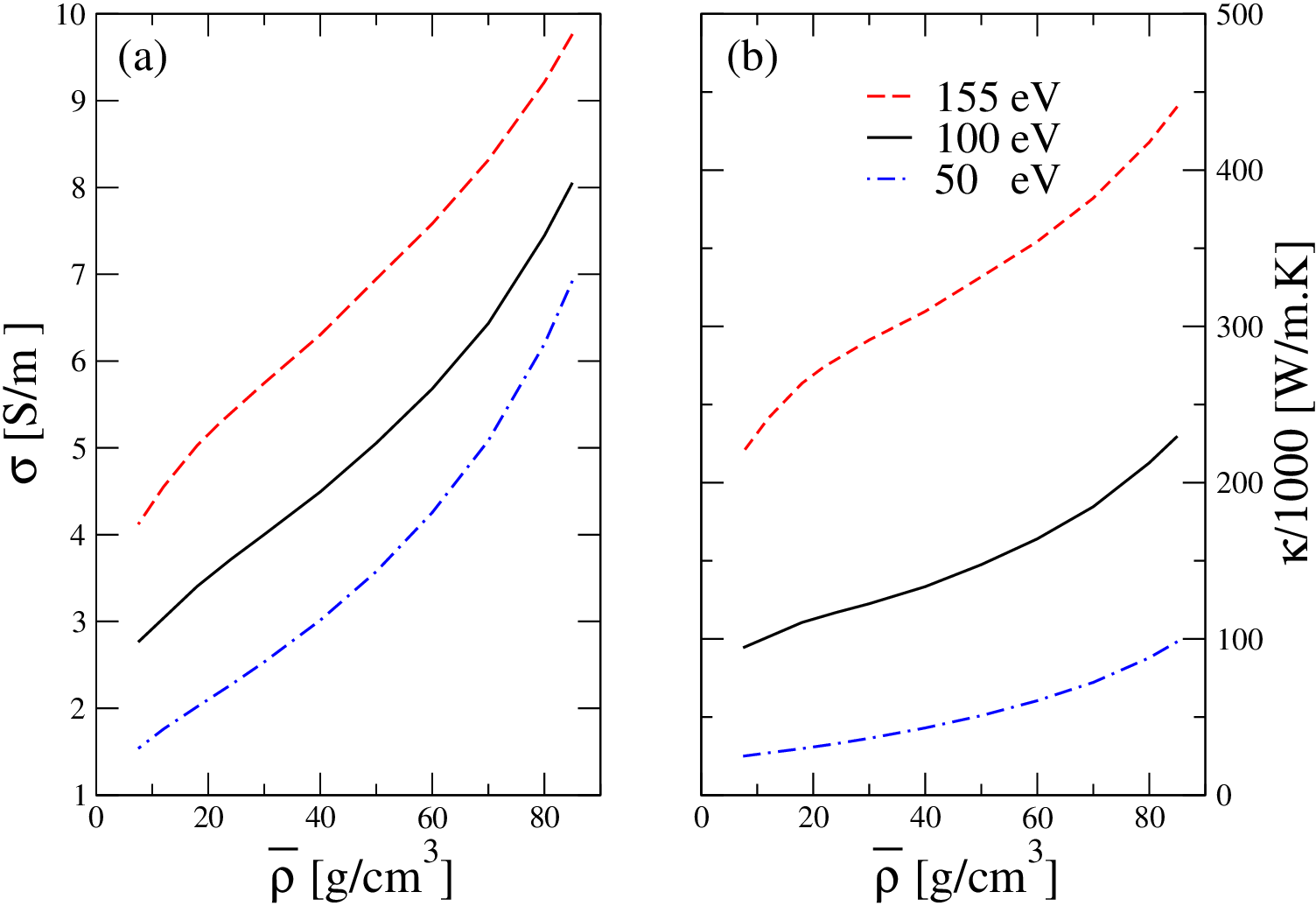}
 \caption{\label{sigkappa.fig}(online color) (a) The electrical conductivity
of Be at 50, 100 and 155 eV, for the range of densities $\bar{\rho}$ in the
range 5-85 g/cm$^3$. 
(b) The thermal conductivity for Be plasmas of panel (a) calculated using
the finite-$T$ Lorenz-number formalism of Ref.~\cite{Ln.24}
}
\end{figure}

The transport properties of Be at the high compressions and temperatures
that appear in the NIF-study are of interest, and involve electron scattering
from the ions, rather than photon scattering. While the coupling of the electrons
to photons is weak and can be treated semi-classically, the scattering of
 electrons from the
electron-ion system involves strong scattering effects, especially since the
internal structure of the hot  Be-ions no longer supports Pauli blocking.
Hence the electrical conductivity $\sigma$ and the thermal conductivity $\kappa$
have to be evaluated via a T-matrix calculation. As $\bar{Z}$ is large,
 electron-electron scattering effects may be neglected~\cite{SpHm53}.
Our calculations for $\sigma$ and $\kappa$ for Be, following the method
 of Ref.~\cite{Ln.24}
are presented in Fig.~\ref{sigkappa.fig} for the three isotherms 50, 100 and
155 eV.

\section{Discussion} We have shown that the neutral-pseudo-atom model, i.e.,
 a rigorous DFT-based average-atom model provides rapid and accurate
interpretation of XRTS measurements, giving results in close agreement with
more numerically demanding and costly methods like PIMC or $N$-atom QMD
 simulations. QMD calculations in the density and temperature range of the
NIF experiment would consume months of computational time, while the NPA and other
AA-methods take only minutes. Here we have shown that the NPA achieves results
comparable to the PIMC and QMD calculations. Most AA codes could achieve similar
or better accuracy by adhering more closely to density-functional theory. 

The introduction of the Rayleigh weight $R_w(k)$, and
the elastic to inelastic ratio $r(k)$ etc., in recent publications~\cite{DornheimRw24}
makes the exploitation of XRTS data via NPA/AA techniques even simpler. Furthermore,
 unlike PIMC or QMD, the scattering wavevector $k$ of the XRTS signal
accessible via NPA is not limited by the ``box-size" of the simulations. 

While NPA can (at least in principle) provide all the structure
 factors $S_{a,b}(k)$, with
$a,b$ = electrons or ions, the use of a classical map for calculating
 the $S_{ee}(k)$ reduces computations to mere minutes, eminently
implementable in ``on-the-fly" calculations within large codes.
Unlike QMD or PIMC, the NPA approach directly yields precisely definable single-ion
properties such as the  mean ionization $\bar{Z}$, the chemical potentials of
ions and electrons, the minimum value of the free energy of the system and all
related physical properties. By confirming the agreement of the NPA predictions 
of the Rayleigh weight etc., (here, via XRTS measurements), we  validate
$\bar{Z}$ and other one-body properties of the average-atom model used in the NPA.
This is merely a corollary of the DFT prescription that a complete one-body
description  in terms of $n(r),\rho(r)$ and their exchange-correlation
functionals exists by virtue of Mermin's extension of the Hohenberg-Kohn theorem.

Unlike PIMC which is limited to low-$Z$ materials and
higher temperatures, the  NPA and classical-map methods are seamlessly applicable
from very low to  high temperatures. An exhaustive
discussion of the equation of state of Be using the NPA was
 given by Perrot in 1993~\cite{Pe-Be}, for compressions up to six,
 and $T$ up to 100 eV. Perrot had also
confirmed the 0-Kelvin isotherm and examined the application
 of the NPA to solid Be. More details of moderate-$T$ states
 of Be,
ionization data up to 155 eV and 85 g/cm$^3$, and comparisons with
QMD calculations are given in the appendix. 

\appendix

\section{}
In this appendix we use the notations, abbreviations and units used in the
main text.
\subsection{Details of the NPA calculation}
\label{npa.sec}
We consider a  sphere of radius $R_c=10 r_{ws}$ at low temperatures (i.e., $T/E_F\le 1$).
This is called the correlation sphere and it is sufficiently large such that
all PDFs $g_{ab}(r)$ have reached unity as $r\to R_c$. This is a significant difference
from many popular AA models which use an ion-sphere of radius $R_c=r_{ws}$ as
the volume of the pseudoatom. The thermal de Broglie length $\lambda_{deB}$
of electrons at temperature $T$ in energy units
 is given by $\lambda_{deB}=1/(2\pi\bar{m_e}T)$, where $\bar{m_e}$ is the effective
mass of the electron.
As $T\to 0$, $\lambda_{deB}$ diverges to large values exceeding $r_{ws}$.
Hence, electrons cannot be meaningfully confined to within an $r\le r_{ws}$ 
except at sufficiently high $T/E_F$. Thus the use of the ion-sphere model
for average atoms is justified only for sufficiently large $T/E_F$. 

The chemical potential $\mu$ used in ion-sphere AA models is determined  by
an equation of the form
\begin{eqnarray}
\label{mu.eqn}
Z_n&=&\int_0^{r_{ws}} 4\pi r^2 n(r) dr \\
n(r)&=&\sum_\nu2(2l+1)|\phi_\nu(r)|^2f(\epsilon_\nu,T)\\
f(\epsilon_\nu,T)&=&1/\left[1+\exp\{(\epsilon_\nu-\mu)/T\}\right].
\end{eqnarray}
Here $\phi_\nu, \epsilon_\nu$, with $\nu=n,l$ or $k,l$ are the bound and free
eigenfunctions and eigenvalues of the spin-unresolved AA calculation,
 and $n(r)$ is the total electron-density pileup inside the
ion sphere. The so-evaluated $\mu$ contains an increasingly
large unphysical non-local potential that confines the free electrons within 
the ion sphere as $T\to 0$. This also implies that the $\bar{Z}$ calculated from the
Friedel sum rule for these ion-sphere AA models may even exceed the nuclear
charge~\cite{Fauss21}, where as it should equal $\bar{Z}\le Z_n$. 

Furthermore, a proper DFT
model should have $\mu$ equal to the non-interacting value $\mu^0$. So,
the inclusion of this type of nonlocal potential in the calculation via a
boundary condition (c.f., Eq.~\ref{mu.eqn}) implies that the Euler-Lagrange
condition for the minimum of the total free energy of the electron-ion system implied by
the Kohn-Sham equation is not met. A variety of ways has been
 proposed~\cite{SternZbar07,Murillo13} for the
evaluation of $\bar{Z}$ from  such AA models. This ambiguity should be viewed as
a weakness in such models rather than in the concept of `mean ionization' as such.

The NPA does not use the ion-sphere, but works with a correlation
sphere with a radius $R_c$ typically 5 to 10 times $r_{ws}$,
 large enough to include all the particle
correlations that exist in the physical system.
Thus the correlation sphere nominally contains about 4000 Be nuclei and 16,000 electrons.
For low-density high-$T$ plasmas,
the correlation sphere becomes similar to the Debye sphere.
For high $T$ implementations, the correlation-sphere radius $R_c$
 may in practice be reduced to about $5 r_{ws}$.  One of the nuclei is at the origin of the
coordinates; the field-nuclei (together with any bound electrons) are replaced
 by their (initially unknown) smoothed average ion-density distribution
$\rho(r)=\bar{\rho}g_{ii}(r)$, each ion carrying a charge $\bar{Z}$ which
is to be determined self-consistently to reach a minimum in the Helmholtz free
energy. Similarly, the electron density
 is $n(r)=\bar{n}g_{ei}(r)$ and normally involves the ``free'' electron distribution
$n_f(r)$ that tends to $\bar{n}$ as $r\to \infty$; i.e., as $r\to R_c$ and not $r_{ws}$.
 The bound electron distribution $n_b(r)$ is short-ranged.
 A simple cavity-like trial $g(r)$ for ions is specified by its
 radius $r_{ws}$, itself to be determined self-consistently with
 $\bar{Z}, n(r)$ and $\rho(r)$~\cite{DWP82,eos95}. 
As the NPA is a DFT model, the electrons are mapped to a non-interacting system
at the interacting density. Hence, the chemical potential of the electrons
is just the non-interacting value $\mu^0$.

 The DFT calculation is based on a minimization of the Helmholtz
 free energy  $F([n],[\rho])$ via the coupled Euler-Lagrange equations for the
electron distribution and the ion distribution.
\begin{eqnarray}
\label{dft.eqn1}
\frac{\delta F([n],\rho])}{\delta n}&=&0,\\
\label{dft.eqn2}
\frac{\delta F([n],\rho])}{\delta \rho}&=&0.
\end{eqnarray}
The first of the above equations reduces to the Kohn-Sham equation for electrons in
the correlation sphere.
 The second reduces to a DFT equation for classical particles. This can be reduced
to a form of the modified hypernetted chain equation or treated via molecular dynamics. 
These equations (\ref{dft.eqn1},\ref{dft.eqn2}) are discussed in some detail
in Ref.`\cite{DWP82,Pe-Be} and in Refs.~\cite{ilciacco93,eos95}. The main approximation
made is the neglect of XC-potentials arising from certain electron-ion XC-functionals
 of the form $F^{ei}_{xc}([n],[\rho])$ in most NPA calculations which retain $F^{ee}_{xc}$
 and $F^{ii}_{xc}$. The latter is highly non-local but an explicit form can be given
in terms of $g_{ii}(r)$, Eq.~3.4 of Ref.~\cite{DWP82}.

How to include $F^{ei}_{xc}$ in NPA calculations (if needed)
 is discussed in Ref.~\cite{Furutani90}. In fact, the main criticism against the
NPA leveled  by Chihara~\cite{ChiharaNPA91} is based on the neglect of
 $F^{ei}_{xc}([n],[\rho])$ type XC-corrections. However, comparisons of PDFs, structure
 factors and other properties of WDM systems calculated from the NPA, with those from
more microscopic methods show that the approximations used in the NPA are satisfactory.

 Furthermore, for systems
near metal-insulator type transitions, the issue of partially localized electrons
(``hopping electrons'') needs to be treated, as in Ref~\cite{ hop1992}.
However, the present study on Be is in the dense metallic region and
hence ``hopping electrons'' are not an issue here. Instead, we have an unusual
example of the ``metallization" of even the 1$s$ shell by delocalization into
neighboring atoms.
 
The converged DFT equations~\ref{dft.eqn1},\ref{dft.eqn2}
 for electrons and ions respectively define a minimum in the total free energy,
 inclusive of all many-body  corrections that are accounted for via
 the appropriate e-e and i-i XC-functionals. Thus a sophisticated Saha equation is
 automatically solved in  evaluating the electron distribution $n(r)=n_b(r)+n_f(r)$,
 the ion distribution $\rho(r)$ and the mean ionization  $\bar{Z}$. 
 At the end of the calculation we have $g_{ii}(r), g_{ei}(r)$,
 corresponding $S_{ii}(k)$ and $S_{ei}(k)$ where the $k\to 0$ limit is accurately
accessible, unlike in QMD and PIMC where the small size $L\sim N^{1/3} $ of the
 simulation cells limit the $k$ range to $k>\pi/L$. 

The total free-energy is used to obtain the pressure, internal
energy, compressibility and other EOS data. These, calculated from
the NPA for Be are given in Ref.~\cite{Pe-Be}, for lower compressions
 ($< 6$) and $T\le 100$ eV.

\subsection{Evaluation of the mean ionization $\bar{Z}$ in the NPA}
\label{zbar.sec}
It has sometimes been claimed that the mean ionization $\bar{Z}$ is not
 a properly defined
quantity as ``it cannot be given as {\it the mean value of a quantum operator}''
{\cite{PironBlenski11,SXHu16, Pain2023, Sharma25}, or that it is a ``a heuristic'' 
quantity~\cite{DornheimPIMC24}. The `temperature' $T$ also has no operator in
quantum mechanics; hence, although the same claim could be used against $T$,
those who reject $\bar{Z}$ continue to retain $T$ in their theories. 
We are in fact {\it not} studying pure quantum systems, but our interest is
 in quantum-{\it statistical} systems where a classical heat bath is attached to
 the quantum system. The
temperature $T$, chemical potential $\mu$, and $\bar{Z}$ appear as
Lagrange multipliers in the quantum-statistical theory used~\cite{DWP82}.
Furthermore, one may invoke extended field-theoretic
formulations of quantum statistical theory, e.g, thermofield dynamics~\cite{Umezawa82},
 where operator formulations are available for quantities like $T,\mu$ and $\bar{Z}$.
However, for the case of $\bar{Z}$, it is possible to give an operator
form even within standard quantum theory using the phase shifts of
plane waves scattering off a pseudoatom. The Friedel sumrule, to be
discussed below is in fact an embodiment of that approach.
%N. Wetteand J.-C. Pain, Phys. Rev. E. {\bf 108}, 015205 (2023)
%S. X. Hu, L. A. Collins, V. N. Gonchorov, J. D. Kress, R. L. McCrory, and S. Skupsky,
%Phjys. Plasmas, {\bf 23}, 042704 (2016), Erratum  First principles investigation of
%ionization and thermal conductivity of polystyrene for inertial confinement fusion applications
%V. Sharma and A. J. White, Phys. Rev. Lett. {\134}, 095102 (2025)
% P. A. Stern, S. B. Hanson, B. G. Wilson, and H. A. Isaacs, High Energy Density Physics, {\bf 3},
% 278, (2007).
%G. Fausssurier, C. Blanchard, and M. Bethkenhagen. Phys. Rev. E {\bf 104}, 025209 (2021).% Thomas Gawne, Sam M. Vinko, and Justin S. Wark. Phys. Rev. E {\bf 103}, L023201 (2024). 

On convergence, NPA calculation satisfies the finite-$T$ Friedel sum rule~\cite{DWP82}
 accurately
and provides the value of $\bar{Z}$ adopted in our calculations. 
The NPA provides the 
 phase shifts $\delta_l(k)$ of the continuum eigenstates,
with energy $\epsilon(k,l)=k^2/2$ which is independent of $l$. The $l$ states are
evaluated up to an $l_{max}\sim 30$ or as appropriate. 
\begin{eqnarray}
\label{friedel.eqn}
\bar{Z}&=&\frac{2}{\pi T}\int_0^\infty dk kf(k)\{1-f(k)\}\hat{X} \\
       \hat{X}&= &\sum_l(2l+1)\hat{\delta}_l(k)\\
f(k)&=&1/\{1+\exp(k^2/2-\mu^0)/T\}
\end{eqnarray}
Noting that $f(k)(1-f(k)$ can be expressed as the derivative of the Fermi function,
Eq.~\ref{friedel.eqn} can be written as the meanvalue $\langle \hat{X} \rangle$ where
both a thermal and a quantum mechanical average are taken over the scattering operator
$\hat{X}$, defined in terms of the phase shifts $\hat{\delta_l}$. Here the
 phase shift $\hat{\delta}_l$
is not just a number but the operator that modifies the scattering amplitude. The phase
shifts are viewed as operators acting on the incoming wavefunctions, modifying their
phases and creating the outgoing wavefunction in each channel.

So, contrary to
 claims that $\bar{Z}$ is a quantity without an operator representation,
inadmissible in quantum theories,
we regard it as a Lagrange multiplier for charge neutrality~\cite{DWP82},
having an operator representation
 and a quantum-statistical meanvalue
 that arises in the DFT
 projection of the $N$-atom density to
 a single average atom. Just as the one-body eigenvalues, eigenfunctions etc.,
of DFT theory refer to properties of the non-interacting Kohn-Sham 
electron system, the $\bar{Z}$ and other properties of the NPA belong to
the non-interacting system of pseudoatoms of DFT theory. This approach consistently
provides the total Helmholtz free energy of the system, and related
finite-$T$ thermodynamic and linear transport properties of WDM systems~\cite{Ln.24}. 

XRTS may be considered as an experimental means of measuring $\bar{Z}$
of matter under extreme conditions, while Langmuir probes can be used
for measuring the free electron density per ion (i.e., $\bar{Z}$) in low-$T$ plasmas.
Fit formulae for $\bar{Z}$ for Be as a function of density
for six isotherms in the range (1 eV to 155 eV) are given in
 sec.~\ref{fitZ.sec}.

\subsection{The atomic structure of Be at 20g/cm$^3$ and 155 eV}
\label{atstr.sec}
As seen in Fig.~\ref{Rw22g155ev.fig}, our calculations suggest that the
NIF-Be sample had a density of 20 g/cm$^3$ and a temperature of 155 eV.
A plot of the NPA bound and free-electron densities obtained
 is displayed in Fig.~\ref{denplot.fig}.

The all-electron Kohn-Sham calculation gives only a $1s$ bound state
 with an energy $\epsilon_{1s}=4.5605$, and a mean radius
 $\langle r \rangle$ = 0.4182 a.u., while the nominal Wigner-Seitz radius
 is 1.064 a.u. Hence, the $1s$-state is mostly contained in
 the WS-sphere. Nevertheless, as seen in Fig.~1(b), at these
compressions, the ion-ion pair-distribution function has a significant
 presence inside the WS-sphere. Using the 1$s$-eigenfunction and $f(\epsilon_{1s})$
we estimate that 0.365 electrons are transferred from the central ion
to the neighboring ions in this case.   

\begin{figure}[t]                    % fig. 5
\includegraphics[width=0.96\columnwidth]{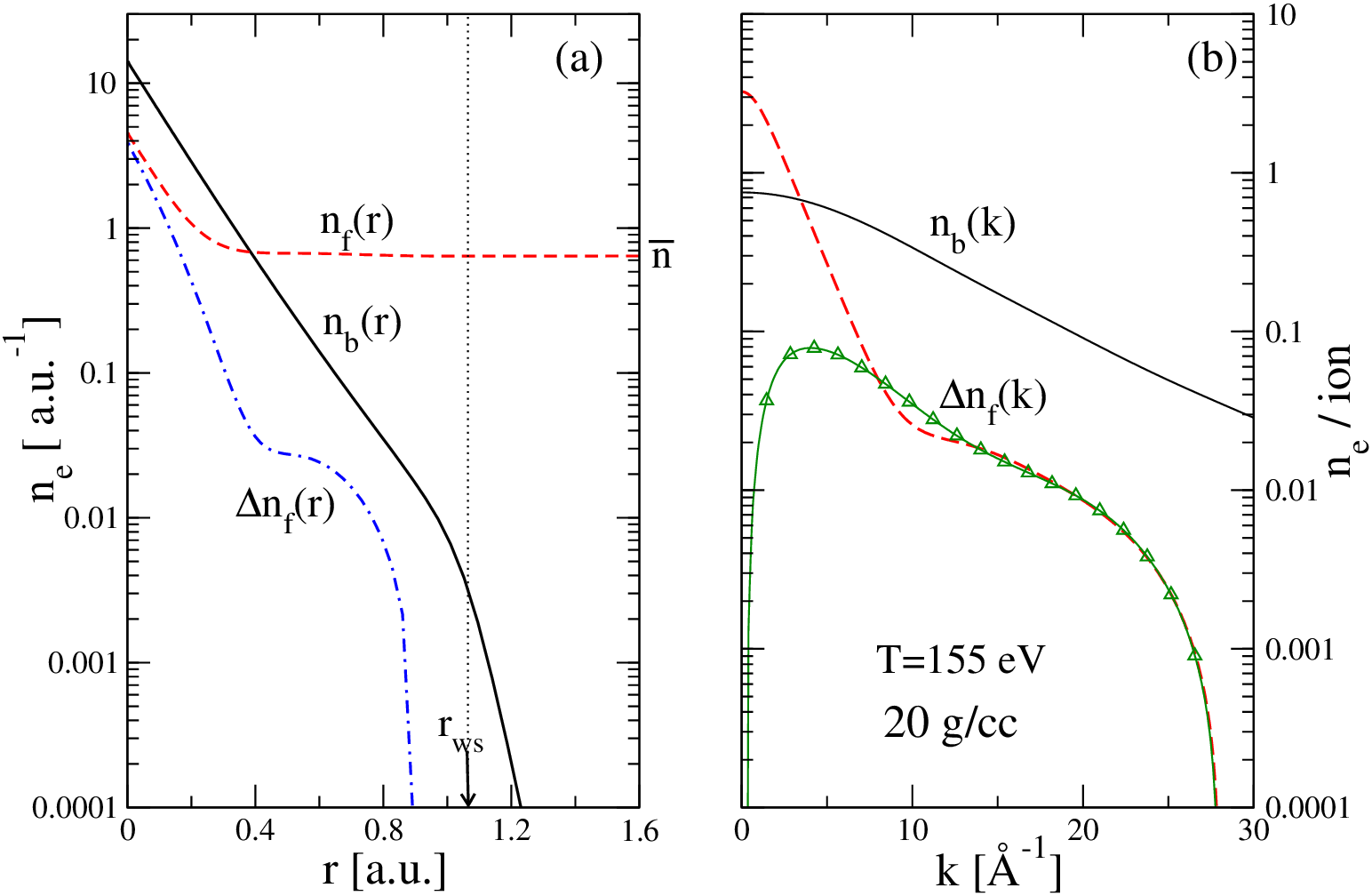}
 \caption{\label{denplot.fig}(online color) (a) The bound and free electron densities
from the NPA calculation. Here $n(r)=n_b(r)+n_f(r)$
 and $\Delta n_f(r)=n_f(r)-\bar{n}$ are shown for Be at 20 g/cm$^3$ and $T$=155 eV.
 Note that the bound-electron profile
extends beyond $r_{ws}$ and hence AA models that  confine electrons to the
Wigner-Seitz cell will differ from our calculations. (b) The
Fourier transforms of these densities are displayed. They are used in calculating
 the ``ion feature'' and the Rayleigh weight of the XRTS profiles.
 The curve with  triangles as data points shows the effect
of a responding ion distribution (approximate two-component model) on the small-$k$ region.}
\end{figure}

\subsection{Fit formulae for the mean ionization $\bar{Z}$}
\label{fitZ.sec}
In Fig.~\ref{zbargud.fig}(a) the mean ionization of Be along four isotherms
has been plotted. Fit formulae for $\bar{Z}$ along
six such isotherms are given below.
\begin{equation}
\label{zcoef.eqn}
\bar{Z}(x)=\frac{(a_0+a_1x+a_2x^2)}{(1.0+b_1x+b_2x^2)}; \; x=\bar{\rho}\;\mbox{g/cm}^3 
\end{equation}
The fit coefficients are given in Table~\ref{zcoefs.tab}. The results for $T=150$, and 155 eV
provide a reasonable estimate of $d\bar{Z}/dT$ near $T=150-155$ eV in parametrized form.

\begin{table}
\caption{\label{zcoefs.tab} Coefficients for $\bar{Z}$ given by fit formula
Eq.~\ref{zcoef.eqn}, for the density range 5 g/cm$^3$-85 g/cm$^3$. For $T$ =1 ev,
and  10 eV, $\bar{Z}$ jumps to 4.0 when $\rho$ exceeds 80 g/cm$^3$. Data on $\bar{Z}$
calculated using the NPA for lower densities and other temperatures, as well
as EOS data  may be found in Ref.~\cite{Pe-Be}.
}
\begin{ruledtabular}
\begin{tabular}{lccccc}
%\begin{tabular}{lllllll}
$T$ [eV] & $a_0$    & $a_1\times 10^2$& $a_2\times 10^3$ & $b_1\times 10$ &  
$b_2\times10^4$ \\
\hline
155      & 3.77176 & 4.28122  & -0.355843 & 0.232129 & -2.03412 \\
150      & 3.83327 & 6.30687  & -0.501881 & 0.312884 & -2.66131 \\
100      & 3.61672 & 11.5306  & -0.835117 & 0.571745 & -4.80523 \\
50       & 2.92788 & 26.3660  & -1.13536  & 1.36315  & -9.57148 \\
10       & 1.98148 & 0.348542 & 0.100605  & 0.0 & 0.0\\
1        & 1.97952 & 0.343568 & 0.126261  & 0.0 & 0 0 \\
\end{tabular}
\end{ruledtabular}
\end{table}

\subsection{The electron-ion pseudopotential and the ion-ion potential}
A key component of these calculations is the construction of a weak $s$-wave electron-ion
pseudopotential $U_{ei}(k)=\Delta n_f(k)/\chi(k)$ that can be used in linear-response theory.
Here $\chi(k)$ is the full interacting e-e response function containing an LFC
consistent with the finite-$T$ XC-potential used for e-e interactions.
The pseudopotential is then used to construct the ion-ion pair-potential
$V_{ii}(k)$ and its Fourier transform $V_{ii}(r)$ in real space. The latter can then be
used in the HNC or MHNC equations, or in an MD simulation to obtain the ion-ion PDF.
The pseudopotential $U_{ei}(k)$ can  be fitted approximately to a Heine-Abarankov form
\begin{equation}
U_{ei}k)=-\frac{4\pi\bar{Z}}{k^2}\left[D\frac{\sin(kr_c)}{kr_c}-(1-D)\cos(kr_c)\right].
\end{equation}
where $k$ is in atomic units. 
We consider the case $\bar{\rho}=20$ g/cm$^3$, and $T=$ 155 eV., $\bar{Z}$ = 3.247
with $k_F$=2.6704/a.u. 
Then, for the range 0 to 2$k_F$,  
the well-depth parameter $D$=0.995835, while the core-radius $r_c=0.690797$. 
As the electron response function $\chi(k)$ drops
rapidly for $k>2k_F$ the above parametrization is sufficient for most purposes.
In our calculations we have, however, used the full numerical tabulation rather than
the fitted forms.

Once the pseudopotential is obtained, the ion-ion pairpotential $V_{ii}(k)$ can
be obtained using linear-response theory. The $r$-space forms of the pair-potential
can be fitted to a Yukawa-Friedel-tail form~\cite{DW-yuk22}. In the present case, a
simple Yukawa form is found to be sufficient for $r/r_{ws}\le 4$, 
with $r_{ws}$ = 1.06427 a.u.
\begin{equation}
\label{potform.eqn}
V_{\rm y}=(a_{\rm y}/r)\exp(-k_{\rm y}r).
\end{equation}
The values of the fit parameters $a_{\rm y},k_{\rm y}$
for the case $\bar{\rho}=20$ g/cm$^3$, and $T=$ 155 eV.,
are 12.8754 and 1.12975 respectively.  
The pair-potential at this high temperature state
is not significantly oscillatory for $r/r_{ws}>4$, behaving nearly as a Yukawa
form with parameters $a_{\rm y}$ = 2.33447, and $k_{\rm y}$=0.710933
 for $4 < r/r_{ws} <8$.

\subsection{The calculation of $S_{ee}(k)$ within NPA and via a classical map}
\label{clasmap.sec}
The NPA calculation yields continuum wavefunctions $\psi_{kl}(r)$ with energies
$\epsilon_{kl}=k^2/2$, and these can be used (instead of plane waves)
 to construct the electron response function
$\chi_{ee}(q,\omega)$~\cite{DW-unpub}. However, we explore below a
simpler approach.

The NPA calculation replaces the many-electron e-e interaction by a
one-body potential usually referred to as the XC-potential, together with the
one-body Hartree potential. In effect, the interacting plasma is replaced
by {\it a Lorentz plasma} where only the electron-ion interaction has to be dealt with,
using the Kohn-Sham equation. The response function $\chi(k)$ of the interacting
electron fluid also simplifies in that the e-e local-field correction
 (LFC) $G_{ee}(k)$ can also be expressed using a density derivative of
 the XC-potential. The Hartree potential reduces to zero for a uniform
 electron fluid (UEF), or for a two-component system (TCF)
of electrons and ions forming a fluid of uniform density. 
\begin{eqnarray}
\label{chi-app.eqn}
\chi_{ee}(k,\omega) &=& \chi^0_{ee}(k,\omega)/D_{ee}(k,\omega),\; \mbox{UEF}  \\
D_{ee}(k,\omega) &=& 1-V_k\{1-G_{ee}(k)\}\chi^0(k\omega)
\end{eqnarray}
As electrons obey quantum mechanics, obtaining the static structure factor $S_{ee}(k)$
usually requires an evaluation of $S^0_{ee}(k,\omega), G_{ee}(k,\omega)$ and an
integration over $\omega$, even with the assumption that the LFC is
 independent of $\omega$.
In the classical limit:
\begin{equation}
S_{ab}(k)=-\frac{T_{cf}}{\sqrt{\bar{n}_a\bar{n}_b}}\chi_{ab}(k,0).
\end{equation}

Hence $\omega$ integrations etc., can be side-stepped by using a
mapping~\cite{prl1} of the electrons to a classical Coulomb fluid, with the
classical-fluid temperature $T_{cf}$ defined by the following equations:
\begin{eqnarray}
\label{tcf-app.eqn}
T_{cf}&=&\surd(T^2+T_q^2) \\
T_q&=&E_F/(1.3251-0.1779\surd{r_s})
\end{eqnarray}
The particles used in the classical map (cm) interact with a potential $\phi_{ij}(r)$
consisting of a Coulomb interaction
supplemented by a Pauli exclusion potential that exactly recovers the Fermi-hole
in the e-e PDFs. The Coulomb interaction itself (represented by the operator $1/r$)
has a diffraction correction due to the quantum nature of the electrons.
The non-interacting PDF, viz., $g^0_{ee}(r)$ and $S^0_{ee}(k)$ are
 exactly known (Eqs. 3-5 of Ref.~\cite{prl1}). 

The
fully interacting $S_{ee}(k)$ is obtained using $\phi_{ij}(r)$ in an HNC equation
or an MD simulation where  $T_{cf}$ is chosen so that the classical fluid
 has exactly the same XC-energy
as the quantum electron fluid. Thus, since the Hartree energy is zero, the classical
fluid and the quantum fluid have the same ground state energy and the same
XC-energy. DFT states that, in that case the charge distributions should be the
equilibrium charge distributions. It is shown in Ref.~\cite{prl1}, and other
publications~\cite{SandipDufty13,LiuWuCHNC14}, that the $g_{ee}(r), S_{ee}(k)$
and the total free energy obtained via this classical-map approach agree
 closely with results obtained by quantum Monte-Carlo and other microscopic methods. 
The CM-potentials, when used with an HNC equation to generate PDFs is referred to as
 the CHNC method. Alternatively MD may be used instead of the HNC equation. The HNC
and MD procedures agree closely because the bridge-corrections needed in the
3D electron-fluid problem tend to be negligible even for very large values of the
coupling parameter $r_s$, unlike in the case of the 2D-electron fluid~\cite{prl2}.

The local-field correction $G_{ee}(k)$ used in the NPA calculations is derived
 from the classical-map procedure via
\begin{equation}
\label{Gee-app.eqn}
G_{ee}(k)=1-(T_{cf}/\bar{n})(1/V_k)\{1/S_{ee}(k)-S^0_{ee}(k)\}
\end{equation}
 Hence the NPA calculation
uses an interacting $S_{ee}(k)$ which is exactly what is also given by the classical-map
procedure. This $S_{ee}(k)$ is displayed in Fig.~3(a) for $\bar{\rho}$=7.5 and
 30 g/cm$^3$ at 155 eV. However, this $S_{ee}(k)$ is the e-e structure factor for
the spin-unpolarized  uniform electron fluid in a non-responding medium. 
This is sufficient for addressing the XRTS signal for scattering wavevectors
 $k_{sc}>k_F$ since the ion-density fluctuations
which are ignored in the $S_{ee}(k)$ of the UEF become relevant only at smaller $k_{sc}$
values. Addressing smaller $k_{sc}$ data requires a calculation of $S_{ee}(k)$ that
explicitly accounts for the two-component (electrons+ions) nature of
the plasma~\cite{Furutani90}. Extending the classical-map approach to electron-ion
systems requires a theory for the effective temperature of the electron-ion interaction.
Currently, this is a subject for further investigation.

\begin{figure}[t]                    % fig 6
\includegraphics[width=0.96\columnwidth]{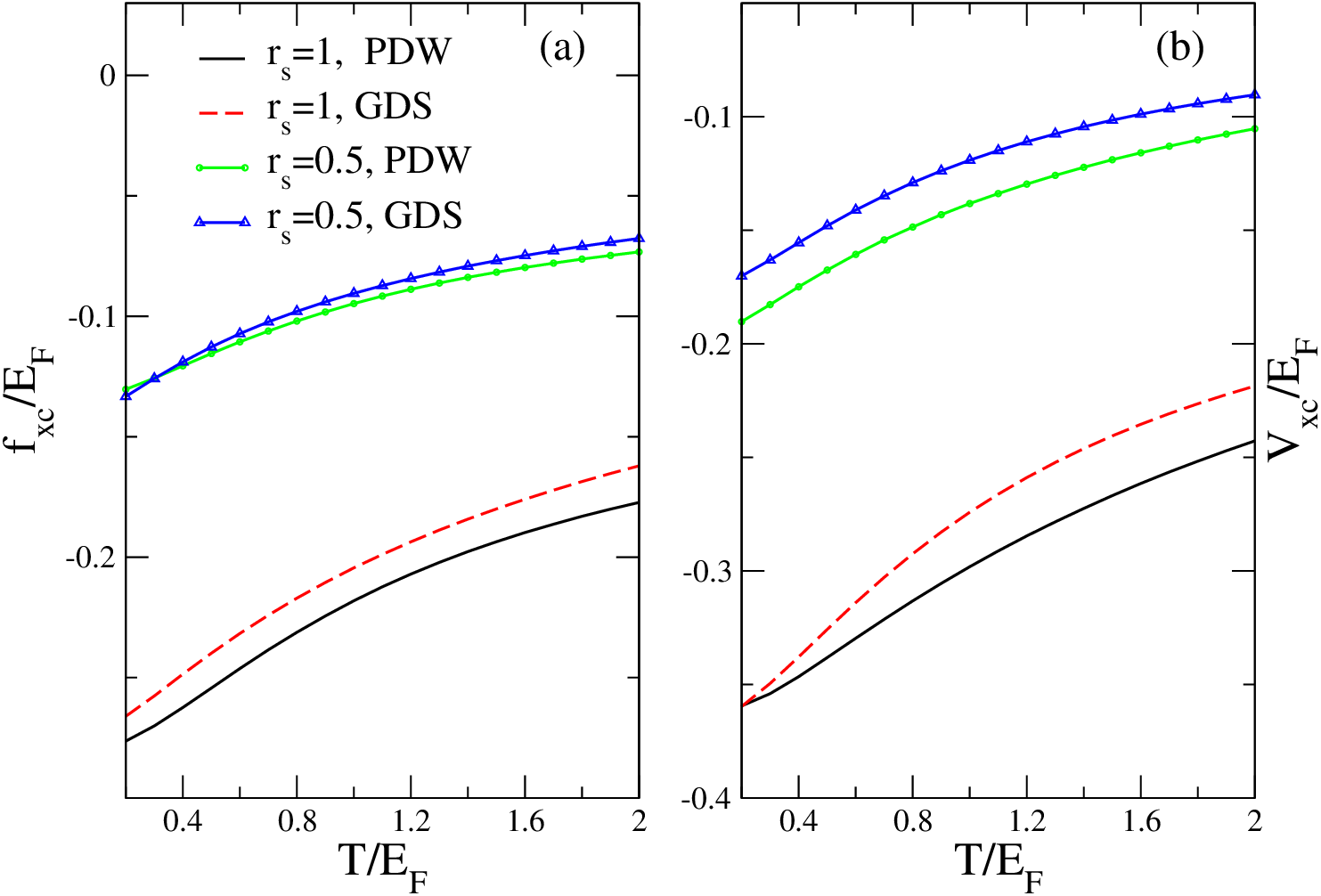}
 \caption{\label{fxc.fig}(online color) (a) A comparison of the free-energy XC
 functionals of Perrot and Dharma-wardana~\cite{PDWXC}, and that of Groth et al
~\cite{GDS17} in the range of densities ($\bar{rho}$) indicated by the $r_s$ parameter,
 and temperature $t=T/E_F$. The $T=0$ functional used in QMD calculations is also shown.
 (b) The XC-potentials $V_{xc}(r_s,t)$ that enter into the Kohn-Sham calculations
are displayed.
}
\end{figure}

\subsection{Choice of finite-$T$ XC-functionals}
\label{xcfunc.sec}
QMD calculations that use $N$-atom DFT, with $N\sim 64-512$ atoms, as implemented in
large codes like the VASP~\cite{VASP} use the $T=0$ functional
and do not implement a  $T$-dependent e-e 
XC-functional. In the case of the NIF-Be, $T=155$ eV and densities are in the range
20$\le \bar{\rho} \le 30$. Hence we have $0.719\le r_s\le$ 0.635 and
 1.59 $\le T/E_F \le$1.25
as the relevant range of $r_s, t=T/E_F$ to be used in e-e XC-calculations. The
 XC-functional of PDW~\cite{PDWXC} for the Helmholtz free energy $F(r_s,t)$
 has been fitted to classical-map calculations within the range $1\le r_s \le 10$, while
 the free energy fit of Groth et al (GDS)~\cite{GDS17} used a wider range of data
 from PIMC and QMC calculations. A comparison of $f_{xc}(r_s,t)$, i.e., the XC-free
 energy per electron, and the corresponding $V_{xc}(r_s,t)$ from the two
 parametrizations covering the region of interest in $r_s,t$
are given in units of $E_F$ in Fig.~\ref{fxc.fig}(a) and (b) respectively.
 It is seen that the two finite-$T$
 $f_{xc}$, and similarly the $V_{xc}$ are in fair agreement. The Be calculations
 reported here (via the NPA code) implements the finite-$T$ PDW functional to
 capture  e-e exchange and correlation effects.

The Kohn-Sham equations do not use the free-energy functional, but use the XC-potential
 defined by:
\begin{equation}
\label{vxc-eqn}
V_{xc}(r_s,t)=\frac{d F_{xc}(r_s,t)}{d n(r_s)}
\end{equation}
Since $F_{xc}(r_s,t)$ is given as a parametrized equation in $r_s,t$, it is convenient to
use an analytical derivative of the $F_{xc}(r_s,t)$ fit. However, the
analytic form of the derivative of the fit-form may not be adequate to
reproduces the XC-potential in all regions of $r_s,t$ although the fit reproduces the
Helmholtz free energy. In fact, it is known that all the available
fitted free-energy functionals, viz., PDW, Karasiev et al, and GDS fail to
 produce physically reasonable specific heats (where two temperature derivatives of the
fit function are invoked) in certain  ranges of $r_s,t$~\cite{Karasiev19}.
 A method of directly evaluating the density
derivative of the XC-potential obtained from the classical-map, without using
 a derivative of the fitted  $F_{xc}(r_s,t)$ is available from Eq.~\ref{Gee-app.eqn}
 for the local field correction to the electron-fluid response
 function. The same possibility exists for PIMC and QMC calculations.
 Hence an additional constraint can be placed on the analytic form
of the $V_{xc}(r_s,T)$ that has to be obtained from the fitted $F_{xc}(r_s,t)$.
 Additionally, the classical-map procedure can probably be contrived to directly
 evaluate $V_{xc}(r_s,t)$. Such developments are left for future work.

\begin{figure}[t]                    % fig 7
\includegraphics[width=0.96\columnwidth]{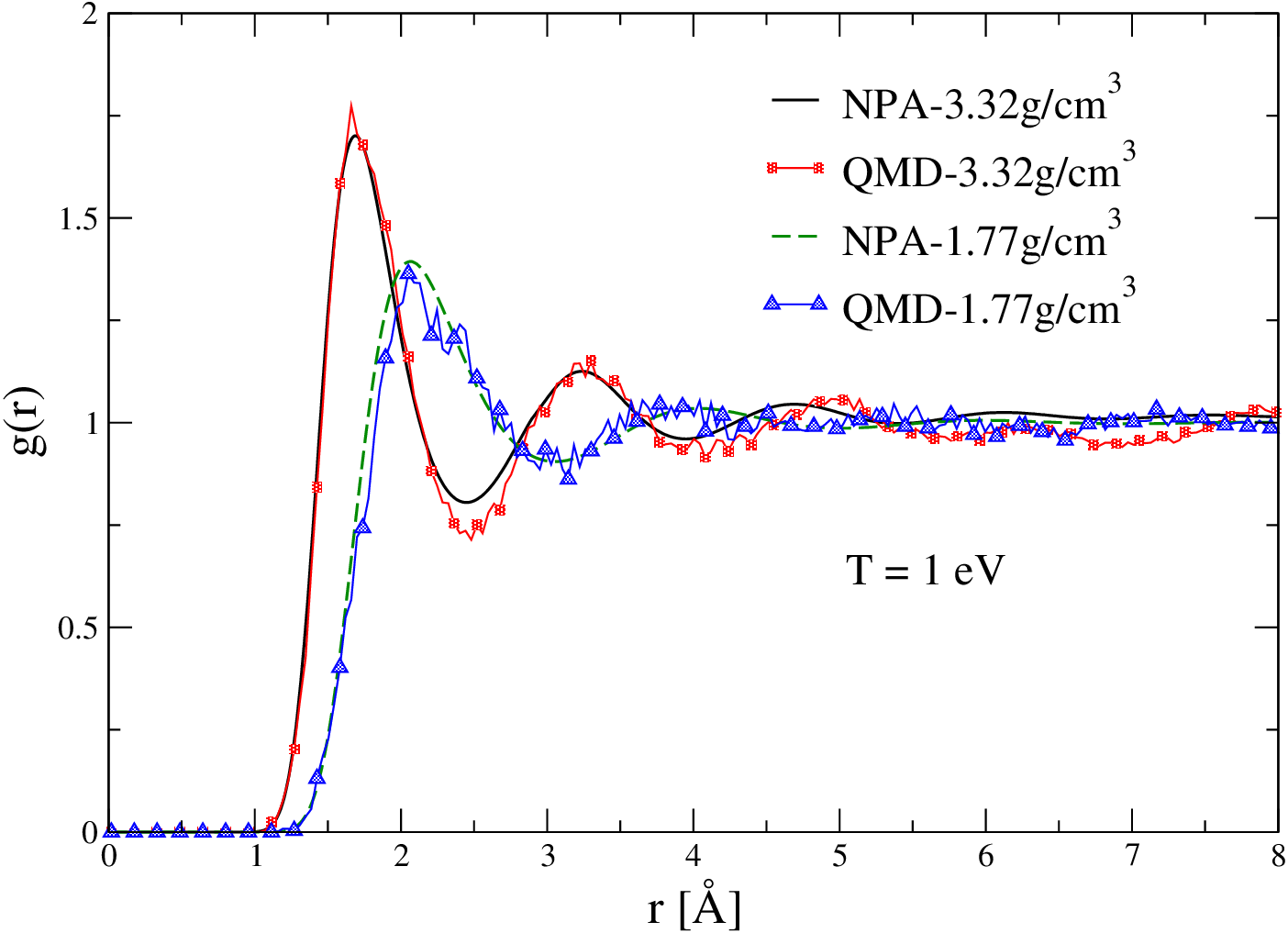}
 \caption{\label{VaspComp.fig}(online color) The Be-Be
$g(r)$ for liquid-Be at $\bar{\rho}$=1.177 and at 3.282 g/cm$^3$ and $T$=1 eV, 
 calculated using the NPA and the MHNC-integral equation, compared with the PDF from QMD
using the Vienna ab-initio simulation package (VASP).
}
\end{figure}

\begin{figure}[t]                    % fig 8
\includegraphics[width=0.96\columnwidth]{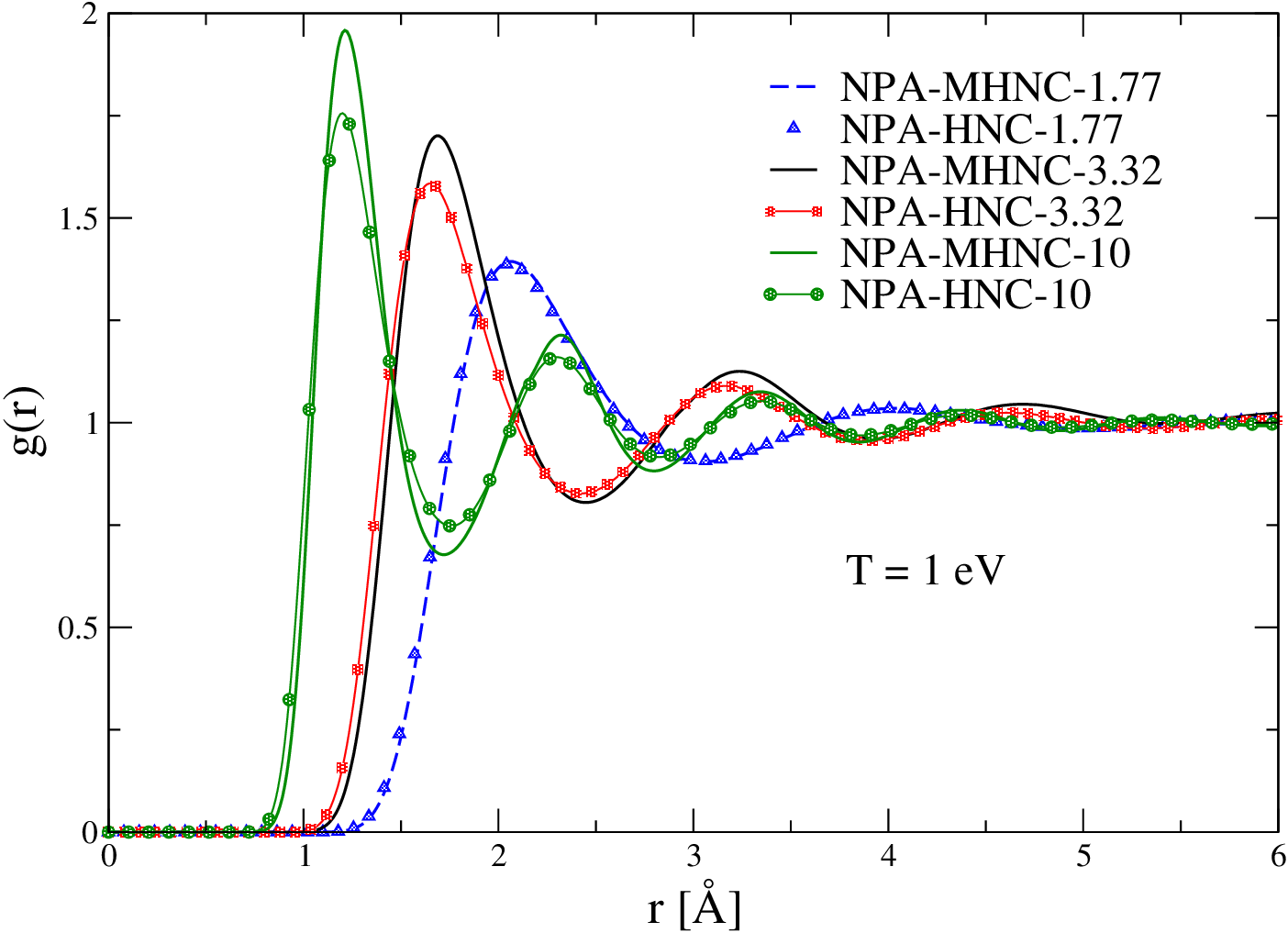}
 \caption{\label{hnc-mhnc.fig}(online color) The Be-Be
$g(r))$ for liquid-Be at $\bar{\rho}$=1.177, 3.282, and 10g/cm$^3$
 and $T$=1 eV, calculated using the NPA ion-ion pair potentials, via the
 HNC and MHNC integral equation, where the latter includes
bridge-diagram contributions modeled using Lado-Foils-Ashcroft
theory~\cite{LFA83}.
}
\end{figure}

\subsection{Comparison of PDFs obtained from NPA and VASP calculations}
In Fig.~1 we compared NPA results with those of PIMC simulations to
establish the extent of agreement that we have between the two methods. 
Here we further authenticate our NPA model for Be calculations
by comparing the PDFs (Fig.~\ref{VaspComp.fig}) obtained for Be
 at 1.117g/cm$^3$, 3.282 g/cm$^3$
at $T$=1 eV with those from a QMD calculation using the VASP~\cite{Vasp-lqdSi,VASP}.
Here the modified Hypernetted-chain (MHNC) equation has been
used to calculated the PDFs using the ion-ion pair potential
obtained from the NPA calculation.
 
Sixty four Be nuclei and 256 electrons were used in an all-electron
QMD simulation using VASP. An all-electron Be-pseudopotential
supplied with the VASP was used.
 An energy cutoff of $\sim$ 260 eV was employed.
The (Monkhorst Pack) $k$-grid was used and calculations
were done at the $\Gamma$ point (0,0,0). Gaussian smearing
(SMEAR=0) with a smearing of 0.1 eV was imposed, as recommended
in Ref.~\cite{Vasp-lqdSi}.
The temperature was set using a Nos\'e-Hoover thermostat.
The Purdew-Burke-Ernzerhof XC-functional~\cite{PBE96} was used for
electrons. It was ensured that the occupation in the highest
 energy state is less
than 0.0001 even at the temperature (1 eV) studied for this comparison.
This establishes the validity of our methods even in the
low-temperature regime where strong-coupling effects are important.

In fig.~\ref{hnc-mhnc.fig} we compare the MHNC and HNC calculations for 
$g(r)$ for three densities at 1 eV to establish the importance of
bridge-graph contributions to ion-ion correlations at these densities at $T$= 1 eV.
The bridge corrections are seen to be negligible for the low-density
weakly-coupled case (1.117g/cm$^3$) studied here. Similarly, bridge corrections
become unimportant for the temperatures relevant to the NIF experiment. 
The hard-sphere packing fraction $\eta$ needed to satisfy the Lado-Foils-Ashcroft
 criterion~\cite{LFA83}
at 1 eV and 10 g/cm$^3$ is 0.43, while at 100 eV it becomes less than 0.03
and may be neglected.


\begin{thebibliography}{99}
\bibitem{Ng05}  
%  Idealized slab plasma approach for the study of warm dense matter
 A. Ng, T. Ao, F. Perrot, M.W.C. Dharma-wardana, M.E. Foord,
Laser and particle beams, {\bf 23}, 527-537 (2005).
%    https://doi.org/10.1017/S0263034605050718

\bibitem{Poole24}
H Poole, M. K. Ginnane, M. Millot, H. M. Bellenbaum, 
G. W. Collins, S. X. Hu,  D. Polsin, R. Saha, J. Topp-Mugglestone,
T. G. White, D. A. Chapman, J. R. Rygg, S. P. Regan,
and G. Gregori.
%Multimessenger measurements of the static structure of shock-compressed liquid silicon at 100 GPa
Physical Review Research {\bf 6}, 023144 (2024).
DOI: 10.1103/PhysRevResearch.6.023144
% hannah.poole@physics.ox.ac.uk

\bibitem{Drake2018}    % ref 3
 R.P. Drake, High-Energy-Density Physics: Foundation of
Inertial Fusion and Experimental Astrophysics, Graduate
Texts in Physics (Springer International Publishing,
2018).


\bibitem{Betti2016}
 R. Betti, O. A. Hurricane, Inertial-confinement fusion with lasers, 
Nature Physics {\bf 12}, 435-448 (2016).

\bibitem{GaffneyHDP18}
J.A. Gaffney, Suxing Hu, P. Arnault..E. Zurek et al
High Energy Density Physics, Aug 2018
https://doi.org/10.1016/j.hedp.2018.08.00



\bibitem{McBride-Si-19}
E. E. McBride, A. Krygier, A. Ehnes, E. Galtier, M. Harmand, Z. Kon\^{o}pkov\'{a},
 H. J. Lee, H.-P. Liermann, B. Nagler, A. Pelka, M. R\"{o}del, A. Schropp, R. F. Smith,
 C. Spindloe, D. Swift, F. Tavella, S. Toleikis, T. Tschentscher, J. S. Wark
 and A. Higginbotham. Nature Phys. {\bf 15}, 89-94 (2019). 
%Phase transition lowering in dynamically compressed silicon
%https://doi.org/10.1038/s41567-018-0290-x

\bibitem{Gregori03}
G. Gregori, S. H. Glenzer, W. Rozmus, R. W. Lee, and O. L. Landen. 
Phys. Rev. E {\bf 67}, 026412 (2003).
%Theoretical model for x-ray scattering as a dense matter probe

\bibitem{GlenRed09}
S. H.  Glenzer and Ronald  Redmer, Rev. Mod. Phys. {\bf 81}, 1625 (2009).
%X-Ray Thomson-Scattering

\bibitem{cdwSi20}
%rb-DSF2018Liquid-liquid Phase Transitions in Silicon
M.W.C. Dharma-wardana, Dennis D. Klug, and Richard C. Remsing
Phys. Rev. Lett. {\bf 125}, 075702 (2020). 
doi: 10.1103/PhysRevLett.125.075702

\bibitem{cdw-SiXRTS24}   ! ref 10
M. W. C. Dharma-wardana, Dennis D. Klug, Hannah Poole and G.gregori.
arXive [cond-mat.mtrl-sci]2408.04173 (2024).
%Ionic structure, Liquid-liquid phase transitions, X-Ray diffraction,
%and X-Ray Thomson scattering in shock-compressed Liquid Silicon
%in the 100-200 GPa regime.



\bibitem{driver12}
K. P. Driver, % {\it et al.,}
and B. Militzer,
 Phys. Rev.  Lett. \textbf{108}, 115502 (2012).

\bibitem{Militzer2015}
 B. Militzer, K. P. Driver, 
 Phys. Rev. Lett. 115, 176403 (2015).
%Development of path integral monte carlo simulations with local-ized nodal surfaces for second-row elements, Phys. Rev. Lett. 115, 176403 (2015).

\bibitem{Harb-DSF2018}   %Ref.13
L. Harbour, G. D. F\"{o}rster, M. W. C. Dharma-wardana, and Laurent J. Lewis.
Phys. Rev. E {\bf 97}, 043210 (2018).
%ion-ion dynamic structure factor, acoustic modes, and equation of state of
% two-temperature warm dense aluminu


\bibitem{Doppner23}
 T. D\"{o}ppner, M. Bethkenhagen, D. Kraus, P. Neumayer,
D. A. Chapman, B. Bachmann, R. A. Baggott, M. P.
B\"{o}hme, L. Divol, R. W. Falcone, L. B. Fletcher, O. L.
Landen, M. J. MacDonald, A. M. Saunders, M. Sch\"{o}rner,
P. A. Sterne, J. Vorberger, B. B. L. Witte, A. Yi, R. Redmer,
 S. H. Glenzer, and D. O. Gericke.
% `Observing the onset of pressure-driven k-shell delocalization', 
Nature {\bf 618}, 270-275 (2023).

\bibitem{SouzaXRTS2014}
A. N. Souza, D. J. Perkins, C. E. Starrett, D. Saumon, and S. B. Hansen.
PhysL Rev. E {\bf 89}, 023108 (2014).
%Predictions of x-ray scattering spectra for warm dense matter


\bibitem{Plage-XRTS15}  % ref. 16
K-U Plageman, H. R. R\"{u}ter, T. Bornath, Mohammed Shihab,
Michael P. Desjarlais, C. Fortmann, S. Glenzer, R. Redmer.
 Phys. Rev. E {\bf 92}, 013103 (2015).
%XRTS Be



\bibitem{xrt-Harb16}
L. Harbour, M. W. C. Dharma-wardana, D. Klug and L. Lewis. 
Physical Review E {\bf 94}, 053211, (2016).

\bibitem{Vorberger24}
J. Vorberger, T. Preston, N. Medvedev, M. P. Bohme, Z. A. Moldabekov, D. Kraus,
and T. Dornheim. Phys. Lett. {\bf 499}, 129362 (2024).

\bibitem{CDW-Pool25}
M. W. C. Dharma-wardana, D. D. Klug, Hannah. Poole, and G. Gregori.
Phys. Rev. E {\bf 111}(1) 015205 (2025).
%Ionic structure, liquid-liquid phase transitions,x-ray diffraction
% and x-ray Thomson scattering in liquid silicon in the 100-200 GPa regime.
DOI:10.1103/PhysRevE.111.015205

\bibitem{BethkenZbar20}
Mandy Bethkenhagen, Bastian B. L. Witte, Maximilian Sch\"{o}rner,
Gerd R\"{o}pke, Tilo D\"{o}ppner,
 Dominik Kraus, Siegfried H. Glenzer, Philip A. Sterne, and Ronald Redmer.
% Carbon ionization at gigabar pressures: An ab initio perspective on astrophysical high-density plasmas.
Phys. Rev. Research {\bf 2}, 023260  (2020).


\bibitem{DornheimPIMC24}   % Ref 20
T. Dornheim, T. D\"{o}ppner, P. Tolias,
M. P. B\"{o}hme, L.B. Fletcher, Th. Gawne, F. R. Graziani,
D. Kraus, M. J. MacDonald, Zh. A. Moldabekov,
S. Schwalbe, D.O. Gericke, and J. Vorberger.
arXive:2402.19113v1 [physics.plasmas-ph]  (2024).
%Unraveling electronic correlations in warm dense quantum plasmas


\bibitem{DornheimRw24}     % Ref 21
T. Dornheim, H. M. Bellenbaum, M. Bethkenhagen, S. B. Hansen,
 M. P. B\"{o}hme, T. D\"{o}ppner, L. B. Fletcher, Th. Gawne,
 D. O. Gericke, S. Hamel,  D. Kraus, M. J. MacDonald,
Zh. A. Moldabekov, Th. R. Preston, R. Redmer, M. Sch\"{o}rner,
 S. Schwalbe, P. Tolias, and J. Vorberger.
arXiv:2409.08591v1 [physics.plasmas-ph]  (2024).

 
\bibitem{DWP82}
M. W. C. Dharma-wardana and F. Perrot. 
Phys. Rev. A {\bf 26}, 2096  (1982).

\bibitem{Pe-Be} %Beryllium
 F. Perrot,  Phys. Rev. E {\bf 47}, 570 (1993).
% Ion-ion interaction and equation of state of a dense plasma: Application to beryllium
% Pe-Be Berillium



\bibitem{eos95}  % Ref 24
F. Perrot and M.W.C. Dharma-wardana,
%Equation of state and transport properties of an interacting multispecies plasma: 
%Application to a multiply ionized Al plasma.
Phys. Rev. E. {\bf 52}, 5352 (1995). 




\bibitem{Vasp-lqdSi}
Details for a standard MD simulation of $l$-Si using VASP are given in
\url{https://www.vasp.at/wiki/index.php/Liquid_Si_-_Standard_MD}.

\bibitem{VASP}
G. Kresse and J. Furthm\"{u}ller, Phys. Rev. B \textbf{54}, 11169 (1996).



\bibitem{prl1}
M. W. C. Dharma-wardana and F. Perrot, Phys. Rev. Lett. {\bf 84}, 959 (2000).
% 3d CHNC, chnc

\bibitem{Bredow15}
R. Bredow, Th. Bornath, W.-D. Kraeft, M.W.C. Dharma-wardana and R. Redmer
Contributions to Plasma Physics, 
 {\bf 55}, 222-229 (2015)
 DOI 10.1002/ctpp.201400080


\bibitem{PDWXC}    %Ref 29
F. Perrot and M. W. C. Dharma-wardana, Phys. Rev. B {\bf 62}, 16536 (2000);
{\it Erratum: } {\bf 67}, 79901 (2003); arXive-1602.04734.
% XC PDW-XC XC-PDW, DWP

\bibitem{Dornheim-Tem23}  % Ref. 30
Tobias Dornheim, Maximilian B\"{o}hme, Dominik Kraus,
Tilo D\"o?ppner, Thomas R. Preston, Zhandos A. Mold-
abekov, and Jan Vorberger,
% ``Accurate temperature di-agnostics for matter under extreme conditions," 
Nature Communications {\bf 13}, 7911 (2022).


\bibitem{Bellenbaum24}  % Ref 31
H. M. Bellenbaum, B. Bachmann, D. Kraus, Th. Gawne, M. P. B\"{o}hme,
T. D\"{o}ppner, L. B. Fletcher, M. J. MacDonald, Zh. A. Moldabcov,
Th. R. Preston, J. Vorberger, and T. Dornheim.
arXiv:2411.06830v1 [physics.plasmas-ph]  (2024).

\bibitem{Chihara2000}
 J. Chihara, J. Phys.: Condens. Matter {\bf12}, 231 (2000).

\bibitem{hop1992}    %Ref. 33
M.W.C. Dharma-wardana and F. Perrot, 
%"Level shifts, continue in lowering,and the mobility edge in dense plasma", 
Phys. Rev. A {\bf 45},5883 (1992).
% hopping electrons

\bibitem{DornheimPhysChemLett24}
T. Dornheim, T. Schoof, S. Groth, A. Filinov, and M. Bonitz.
J. Chem. Phys. {\bf 143}, 204101 (2024)

\bibitem{Kohn86}
W. Kohn, Phys. Rev. B {\bf 33}, 4331 (1986) 
%disc. in XZC functionals

\bibitem{cdw-Carbon10E6-21}
M.W.C. Dharma-wardana, 
%Ionization of carbon at 10-100 times the diamond density and in 
%the 10$^6$ K temperature range.
Phys. Rev. E {\bf 104}, 015201 (2021).

\bibitem{Fauss21}  % Ref.36
%G\'{e}rald Faussurier, Christophe Blancard, and Mandy Bethkenhagen
G. Faussurier, C. Blancard, \&  M. Bethkenhagen.
%Carbon ionization from a quantum average-atom model up to gigabar pressures
Phys. Rev. E {\bf 104}, 025209 (2021).

\bibitem{PDW-Dyson84}
F. Perrot and M. W. C. Dharma-wardana, Phys. Rev. A {\bf 29}, 1378 (1984).
% Dyson equation, beyond DFT


\bibitem{Gawne24}
Thomas Gawne, Sam M. Vinko, and Justin S. Wark. Phys. Rev. E {\bf 103}, L023201 (2024).
%quantifying zbar.

\bibitem{Sharma25}   % Ref. 39
V. Sharma and A. J. White, Phys. Rev. Lett. {\bf 134}, 095102 (2025).
%Born charge model for zbar


\bibitem{Karas18}
%Nonempirical Semilocal Free-Energy Density Functional for Matter under
% Extreme Conditions
%February 2018 Physical Review Letters 120(7)
%DOI: 10.1103/PhysRevLett.120.076401
Valentin V. Karasiev, James Dufty, S. B. Trickey,
Phys. Rev. Lett. {\bf 120}, 076401  (2018).


\bibitem{GDS17}   % Ref.41
S. Groth, T. Dornheim, T. Sjostrom, F.D. Malone, W. Foulkes, M. Bonitz,
%Ab initio exchange-correlation free energy of the uniform electron gas at warm
%dense matter conditions, 
Phys. Rev. Lett. {\bf 119} (13)  135001 (2017).
%http://dx.doi.org/10.1103/PhysRevLett.119.135001.

\bibitem{PBE96}
J. P. Perdew, K. Burke, and M. Ernzerhof, Phys. Rev.
Lett. {\bf 77}, 3865 (1996).


\bibitem{DW-unpub}   % Ref. 43
M. W. C. Dharma-wardana, unpublished.




\bibitem{Furutani90}
 F. Perrot, Y. Furutani and M.W.C. Dharma-wardana,
Phys. Rev. A {\bf 41}, 1096-1104 (1990).
%article stored in papers/DFT17, e-i XC functional, electron-ion corr


\bibitem{ELR98}
M. W. C. Dharma-wardana, nd Fran\c{c}ois Perrot, Phys. Rev. E {\bf 58}, 3705 (1998).
% ELR 1st paper, elr


\bibitem{SpHm53}
L. Spitzer and R. H\"{a}rm, Phys. Rev. {\bf 89}, 977 (1953).


\bibitem{Ln.24}   % Ref. 47
M. W. C. Dharma-wardana (unpublished)
preprint: https://arxiv.org/abs/2404.19692




\bibitem{SternZbar07} 
P.A. Sterne, S.B. Hansen, B.G. Wilson, W.A. Isaacs, 
High Energy Density Phys. {\bf 3}, 278 (2007).

\bibitem{Murillo13}
M. S. Murillo, %{\it et al.,}      % PRE version
 J. Weisheit, S. B. Hansen, and M. W. C. Dharma-wardana,
Phys. Rev. E {\bf 87}, 063113 (2013).


\bibitem{ilciacco93}   % Ref.50
E. K. U. Gross, and R. M. Dreizler,
{\it Density Functional Theory},
 NATO ASI series, {\bf 337}, 625
 Plenum Press, New York (1993).

\bibitem{ChiharaNPA91}
J. Chihara, Phys. Rev. A {\bf 41}, 1247 (1991)

\bibitem{PironBlenski11}
R. Piron and T. Blenski, Phys. Rev. E {\bf 83}, 026403 (2011).
%Variational-average-atom-in-quantum-plasmas (VAAQP) code and virial theorem: 
%Equation-of-state and shock-Hugoniot calculations for warm dense Al, Fe, Cu, and P

\bibitem{SXHu16}
S. X. Hu, L. A. Collins, V. N. Goncharov, J. D. Kress, R. L. McCrory, S. Skupsky.
Physics of Plasmas, {\bf 23}, 042704 (2016).
%First principles investigation of ionizatio and thermal cond. of polystyrene for ICF 
% applications.

\bibitem{Pain2023}
Nadine Wette and J-C Pain, Phys. Rev. E {\bf 108}, 015205 (2023).
%Average-Atom resitivity calculations in expanded metallic plasmas:
%Effect of mean ionization definition.


\bibitem{Umezawa82}
H. Umezawa, H. Matsumoto, and M. Tachiki. {\it Thermo field dynamics and condensed
states}, North-Holland, Amsterdam (1982). 





\bibitem{DW-yuk22}
%Yukawa-Friedel-tail pair potentials for warm dense matter applications
M. W. C. Dharma-wardana, Lucas J. Stanek, and Michael S. Murillo
Phys. Rev. E {\bf 106}, 065208 (2022).
% Published 22 December 2022

\bibitem{SandipDufty13}
J. Dufty, and Sandipan Datta.  Phys. Rev. E {\bf 87}, 032101 (2013) % classical map 




\bibitem{LiuWuCHNC14}
Yu Liu and Jianzhong Wu,
J. Chem. Phys {\bf 141} 064115 (2014).
%An improved classical mapping method for homogeneous electron gases at finite
%temperature, chnc

\bibitem{prl2}
Fran\c{c}ois Perrot and M. W. C. Dharma-wardana,  Phys. Rev. Lett. {\bf 87},
 206404 (2001).
% prl2 for 2D electrons

\bibitem{Karasiev19}
V. Karasiev, S. B. Trickey and J. W. Dufty.  Phys. Rev. B {\bf 99}, 195134 (2019).
% Sp heat

\bibitem{LFA83}
F. Lado, S. M. Foiles and N. W. Ashcroft,  Phys. Rev. {\bf A 26}, 2374 (1983).



\end{thebibliography}
\end{document}